\documentclass[%
twocolumn,
superscriptaddress,
showpacs,
aps,
pra,
floatfix,
]{revtex4-1}
\usepackage{amsmath,amsthm,amsfonts,amssymb,bm}
\usepackage{graphicx}
\usepackage{dcolumn}
\usepackage{bm}
\usepackage{bbm}
\usepackage[colorlinks={true}]{hyperref}
\hypersetup{citecolor={blue}, filecolor={blue}, linkcolor={blue}, urlcolor={blue}}
\usepackage{natbib}
\usepackage{subfigure}
\usepackage{epsf}
\usepackage{color}

\begin{document}

\title{A quantum heat engine with coupled superconducting resonators}
\author{Ali \"{U}.~C.~Hardal}
\affiliation{Department of Physics, Ko\c{c} University, Sar{\i}yer, \.Istanbul, 34450, Turkey}
\affiliation{Department of Photonics Engineering, Technical University of Denmark, \O rsteds Plads 343, DK-2800 Kgs. Lyngby, Denmark}
\author{Nur Aslan}
\affiliation{Department of Physics, Ko\c{c} University, Sar{\i}yer, \.Istanbul, 34450, Turkey}
\author{C.~M.~Wilson}
\affiliation{Institute of Quantum Computing and Electrical and Computer Engineering, 
University of Waterloo, Waterloo, Ontario, Canada N2L 3G1}
\author{\"{O}zg\"{u}r E.~M\"{u}stecapl{\i}o\u{g}lu}
\email{omustecap@ku.edu.tr}
\affiliation{Department of Physics, Ko\c{c} University, Sar{\i}yer, \.Istanbul, 34450, Turkey}
\begin{abstract}
We propose a quantum heat engine composed of two superconducting transmission line resonators interacting with each other via
an optomechanical-like coupling. One resonator is periodically excited by a thermal pump. The incoherently driven resonator induces coherent 
oscillations in the other one due to the coupling. A limit cycle, indicating finite power output, emerges in the 
thermodynamical phase space. 
The system implements an all-electrical analog of a photonic piston. 
Instead of mechanical motion, the power output is obtained as a coherent electrical charging in our case. 
We explore the differences between the quantum and classical descriptions of our system by solving the quantum master equation and classical 
Langevin equations. Specifically, we calculate the mean number of excitations, second-order coherence, as 
well as the entropy, temperature, power and mean energy to reveal the signatures of quantum behavior in the statistical and thermodynamic
properties of the system. We find evidence of a quantum enhancement in the power output of the engine at low temperatures. 
\end{abstract}
\maketitle
\section{Introduction}\label{sec:intro}
Heat engines with quantum working substances, so-called quantum heat engines (QHEs), have attracted much attention recently~\cite{kieu2004second,quan_quantum_2007,quan_quantum_2005,abah_single-ion_2012,
bergenfeldt2014hybrid,henrich_driven_2007,uzdin_multilevel_2014,scully2003extracting,scully2011quantum,
scully2010quantum,brunner_virtual_2012,manzano_entropy_2016,song_one_2016,altintas2015rabi,
ivanchenko_quantum_2015,altintas2014quantum,dag_multiatom_2016,turkpencce2016quantum,
hardal2015superradiant,campisi_power_2016,kosloff_quantum_2014,tonner_autonomous_2005,
anders_thermodynamics_2013,zhang_quantum_2014,zhang_theory_2014,zhang_quantum_2017,mari_quantum_2015,
gelbwaser-klimovsky_work_2015,rosnagel_nanoscale_2014,allahverdyan_extraction_2000,korzekwa_extraction_2016,
perarnau-llobet_extractable_2015,plastina_irreversible_2014,campo_more_2014,roulet_autonomous_2017,hofer2016quantum,hofer2017quantum,hofer2016autonomous,karimi2016otto}. The steam driven mechanical piston 
is an archetype of classical heat engines. Quantum analogs of piston engines have been proposed using optomechanical models, 
where steam is replaced by a photonic gas~\cite{mari_quantum_2015,gelbwaser-klimovsky_work_2015,rosnagel_nanoscale_2014,zhang_quantum_2014,zhang_theory_2014,zhang_quantum_2017}. 
A single-atom piston engine described by an effective optomechanical model has been demonstrated in the classical regime
very recently~\cite{rosnagel_single-atom_2016}. 
The benefits of ``quantumness"~\cite{uzdin_equivalence_2015,mukherjee_speed_2016} as well as the 
quantum-to-classical transition~\cite{quan_quantum-classical_2006} in heat engines are fundamental
problems in the emerging field of quantum thermodynamics. We propose a quantum heat engine composed of a pair of 
superconducting resonators interacting via an effective optomechanical coupling~\cite{johansson_optomechanical-like_2014}. 
It offers an on-chip circuit analog of a piston engine and could be used to explore fundamental quantum properties in 
heat engines. 

\begin{figure*}[!t]
            \includegraphics[width=10cm]{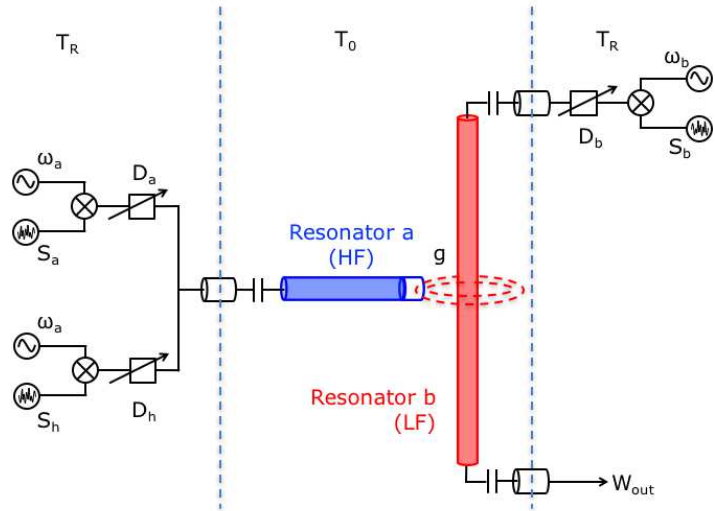}
                \caption{(Colour online) Circuit diagram of the quantum heat engine composed of two superconducting 
                microwave resonators  interacting
                via an effective optomechanical like coupling with strength $g$. Resonators indicated 
                by ``a'' and ``b'', have high frequency (HF) and low frequency (LF) modes with $\omega_a\gg\omega_b$. HF mode is 
                periodically driven by 
                a variable amplitude attenuation ($D_h$) white noise source ($S_h$), corresponding to an effective hot bath at $T_h$. 
                The resonators are on a chip at a background temperature $T_0$. External noise drives are 
               at room temperature $T_R$. Both the LF and HF resonators are subject to additional quasi-thermal sources 
               implemented by amplitude controlled ($D_a,D_b)$ white noise drives $(S_a,S_b)$, which are constantly applied to the corresponding
               resonators.
                }
   \label{fig:fig1}
\end{figure*}

We specifically consider a system of two coupled superconducting resonators~\cite{johansson_optomechanical-like_2014}, 
as an alternative embodiment of the piston cycle used in the single-atom heat engine~\cite{rosnagel_single-atom_2016}. In our case, 
which is shown in Fig.~\ref{fig:fig1}, the resonator modes play the role of vibrational modes of the trapped atom. 
One of the resonators is periodically
driven by a quasi-thermal pump and interacts with the other resonator through
an effective optomechanical coupling~\cite{johansson_optomechanical-like_2014}. The emergence of coherence in one mode by 
incoherent excitation of the other 
is a typical feature of quantum piston engines exploiting optomechanical coupling~\cite{mari_quantum_2015}.
We investigate the quantum statistics of the resonator modes by calculating the mean number dynamics, occupation probability distributions, and second-order correlation functions. We verify that the incoherently driven mode remains thermal while the other
becomes almost coherent. In addition, thermodynamic properties of the system are 
examined by calculating the mean energy versus effective electrical length of the driven resonator and the temperature versus entropy diagrams of the engine. We identify that the
system undergoes an Otto engine cycle.
We find that, after a transient regime, a limit cycle emerges 
in the thermodynamical phase space, indicating finite power output, at the same time that the driven mode induces steady coherent oscillations 
in the other mode. 

Further, we compare the quantum mechanical results with those obtained from the classical Langevin equations of motion. We find significant 
qualitative and quantitative differences between the classical and 
quantum descriptions of the system. In particular, the power output from the quantum piston engine is greater than 
that of the classical one at low temperatures. We argue that the quantum enhancement in power output 
can be observed in superconducting resonators. We explain the effect by identifying the type of quantum correlation 
behind the power output as the
so-called  signal-meter correlation~\cite{holland1990nonideal,mancini1997ponderomotive}
 and compare it with the classical correlations. It is not a universal fact that quantum correlations
can always increase the power output of a piston engine. Recent studies reported exactly the
opposite conclusions where quantum fluctuations cause less power output in a rotor type piston engine~\cite{roulet_autonomous_2017}.

This paper is organized as follows. In Section~\ref{sec:quantumModel}, we introduce the quantum optomechanical engine model based upon
coupled superconducting transmission line resonators.
 In Section~\ref{sec:quantumResults}, we describe our numerical methods and present the results of quantum dynamical simulations. 
 The corresponding classical engine model is introduced in Sec.~\ref{sec:classicalModel} and the results of the classical Langevin equation simulations are given in Sec.~\ref{sec:classicalResults}. These results are compared to the quantum description. In Sec.~\ref{sec:classicalVSquantum},
we discuss the quantum character of the model system in terms of the quantum coherence and correlations.
We conclude in Section~\ref{sec:conc}. 
\section{The Quantum Model}
\label{sec:quantumModel}
Our model system consists of two superconducting transmission line resonators on a chip, held at temperature
$T_{0}$, as shown in Fig.~\ref{fig:fig1}.  A resonator, indicated by ``a'' in the figure, is terminated by a SQUID,
which collects the flux generated by the other resonator, indicated by ``b'', so that an effective optomechanical-like coupling, denoted by $g$, 
between the resonators can be engineered~\cite{johansson_optomechanical-like_2014}. We assume resonator ``a''
with frequency $\omega_a$ is shorter than resonator ``b'' with frequency $\omega_b$ such that $\omega_a\gg\omega_b$.
The high frequency (HF) and low frequency (LF) 
modes of resonator ``a'' and resonator ``b'' can be considered as analogs of the optical and mechanical modes of an optomechanical system, respectively.  The effective 
optomechanical-like coupling between the resonators is expressed as (we take $\hbar=1$)~\cite{johansson_optomechanical-like_2014}
\begin{equation}\label{eq:model}
\hat{H}_{\text{sys}}=\omega_{a}\hat{a}^{\dagger}\hat{a}+\omega_{\text b}\hat{b}^{\dagger}b-g\hat{a}^{\dagger}\hat{a}(\hat{b}+\hat{b}^{\dagger}),
\end{equation}
where $\hat{a}$~($\hat{a}^{\dagger}$) and $\hat{b}$~($\hat{b}^{\dagger}$) are the annihilation (creation) operators for the HF and LF 
modes, respectively. 

In addition to the cold environment at $T_0$, we assume three microwave white noise drives are applied to the resonators.  
These drives are produced by external sources at room temperature{\color{red},} $T_R$. We consider two continuously applied noises
with power spectral densities $S_a$ and $S_b$ on the resonators a and b, respectively. Their amplitudes can
be controlled by variable amplitude attenuators $D_a$ and $D_b$. The HF resonator is subject to another amplitude controlled ($D_h$)
and periodic white noise source with a power spectral density $S_h$. The power spectral densities are assumed to be narrow band, 
centered at the corresponding resonator frequencies $\omega_a$ and $\omega_b$, but much wider than the bandwidth of the
HF resonator. Accordingly, each externally applied 
noise source approximates a one-dimensional black body (thermal) spectrum~\cite{fink_quantum-classical_2010} at 
effective temperatures that can be determined from the Planck distribution functions,
(we take $k_B=1$),
\begin{eqnarray}
\bar{n}_{a}&=&\frac{1}{\exp{(\omega_{a}/T_{a})}-1},\\
\bar{n}_{b}&=&\frac{1}{\exp{(\omega_{b}/T_{b})}-1},\\
\bar{n}_{h}&=&\frac{1}{\exp{(\omega_{a}/T_{h})}-1},
\end{eqnarray}
where $T_h,T_b$ and $T_a$ are the effective temperatures corresponding to the periodic drive, continuous drive on the HF resonator, and 
the continuous drive on the LF resonator, respectively. We assume the
periodic drive is used to engineer an effective hot bath such that $T_h>T_a,T_b$. The mean number of excitations in the cold baths are 
denoted by $\bar{n}_{a}$ and $\bar{n}_{b}$, for the HF and LF modes, respectively. The mean number of excitations in the
periodically modulated hot reservoir is denoted by $\bar{n}_\text{h}$. The one dimensional Planck's law gives the power
spectral densities as $S_x=\omega_x\bar{n}_x$ with $x=a,b,h$.  We consider engineering the two additional cold baths to get more 
flexibility to reach desired steady states in the engine operation, which
may not be achieved in the case of a common single environment at $T_0$. A similar strategy was employed for the case 
of the single-atom piston engine by using an additional cooling laser~\cite{rosnagel_single-atom_2016}. 

The dynamics of the density matrix $\hat\rho$ of the resonator pair can be determined by a master 
equation~\cite{fink_quantum-classical_2010},
\begin{eqnarray}\label{eq:master}
\dot{\hat{\rho}}&=&-i[\hat{H}_{\text{sys}},\hat{\rho}] \\ \nonumber
&+&\kappa_{a}(\bar{n}_{a}+1)D[\hat{a}]+\kappa_{a}\bar{n}_{a}D[\hat{a}^{\dagger}]\\ \nonumber 
&+&\kappa_{b}(\bar{n}_{b}+1)D[\hat{b}]+\kappa_{b}\bar{n}_{b}D[\hat{b}^{\dagger}]\\ \nonumber 
&+&\kappa_{\text{h}}(t)(\bar{n}_{\text{h}}+1)D[\hat{a}]+\kappa_{h}(t)\bar{n}_{\text{h}}D[\hat{a}^{\dagger}]\\ \nonumber 
&+&\kappa_{0}^a(\bar{n}_{0}^a+1)D[\hat{a}]+\kappa_{0}^a\bar{n}_{0}^aD[\hat{a}^{\dagger}]\\ \nonumber 
&+&\kappa_{0}^b(\bar{n}_{0}^b+1)D[\hat{b}]+\kappa_{0}^b\bar{n}_{0}^bD[\hat{b}^{\dagger}]
\end{eqnarray}
Here, $D[\hat{\alpha}]:=(1/2)(2\hat{\alpha}\hat{\rho}\hat{\alpha}^{\dagger}-\hat{\alpha}^{\dagger}\hat{\alpha}\hat{\rho}-\hat{\rho}\hat{\alpha}^{\dagger}\hat{\alpha})$ 
refers to the Lindblad dissipator superoperators with $\hat{\alpha}=\hat{a},\hat{b}$. $\kappa_{a}$ and $\kappa_{b}$ are effective coupling 
constants of the HF and LF modes with their local cold baths, respectively. The coupling coefficient of the HF mode with the effective 
hot bath is denoted by $\kappa_\text{h}(t)$, which has a periodic time dependence. Small background field excitations are denoted by 
$n_0^a$ and $n_0^b$ in the corresponding terms of the master equation 
describing the coupling of the environment at $T_0$ with the HF and LF resonators at rates 
$\kappa_{0}^a$ and $\kappa_{0}^a$, respectively. 

We assume a special case that 
each local cold bath, engineered with the quasi-thermal noise drive, has the same excitation number so that we can introduce
$\bar n_c:=\bar n_a=\bar n_b$, which is possible for $T_a / T_b = \omega_a / \omega_b$. Accordingly we have $T_a\gg T_b$. It may be worth emphasizing that the local  temperatures are ``effective'' and can be 
high (e.g. $100$ K or more) because very little of the noise power is absorbed by the resonators~\cite{fink_quantum-classical_2010}.

The master equation we consider assumes the usual Born-Markov approximations under the weak coupling with the noise sources. Moreover, as we use local effective reservoirs in the dissipators, the dynamics may not be consistent with the second law of thermodynamics at all parameter regimes~\cite{levy_local_2014}.
It is usually assumed that such a local master equation should be reliable when the coupling between the subsystems (here the HF and LF resonators) is sufficiently weak~\cite{hofer2017markovian,gonzalez2017testing}. To determine the validity regime for our case, 
we also used
a ``global" master equation (see Appendix~\ref{app:q-dynamics}) derived for arbitrary optomechanical coupling strengths~\cite{hu_quantum_2015} to compare the results with the ``local" master equation. We found
that both local and global master equation results agree well in the range under consideration $g\le\omega_b$. The regime of $g=\omega_b$ 
is not exactly the ultrastrong coupling regime
of optomechanics, as we have large $\kappa_a>g$. In the temperature regimes we consider, the excitation
of the HF resonator is weak $\langle n_a\rangle < 1$. We explore a regime of single ``photon" optomechanics
which is not well charted.
\section{Quantum Dynamics of the Engine}\label{sec:quantumResults}
\begin{figure}[!t]
	\centering
	\begin{center}
		\subfigure[]{
			\label{fig:fig2a}
			\includegraphics[width=7.5cm]{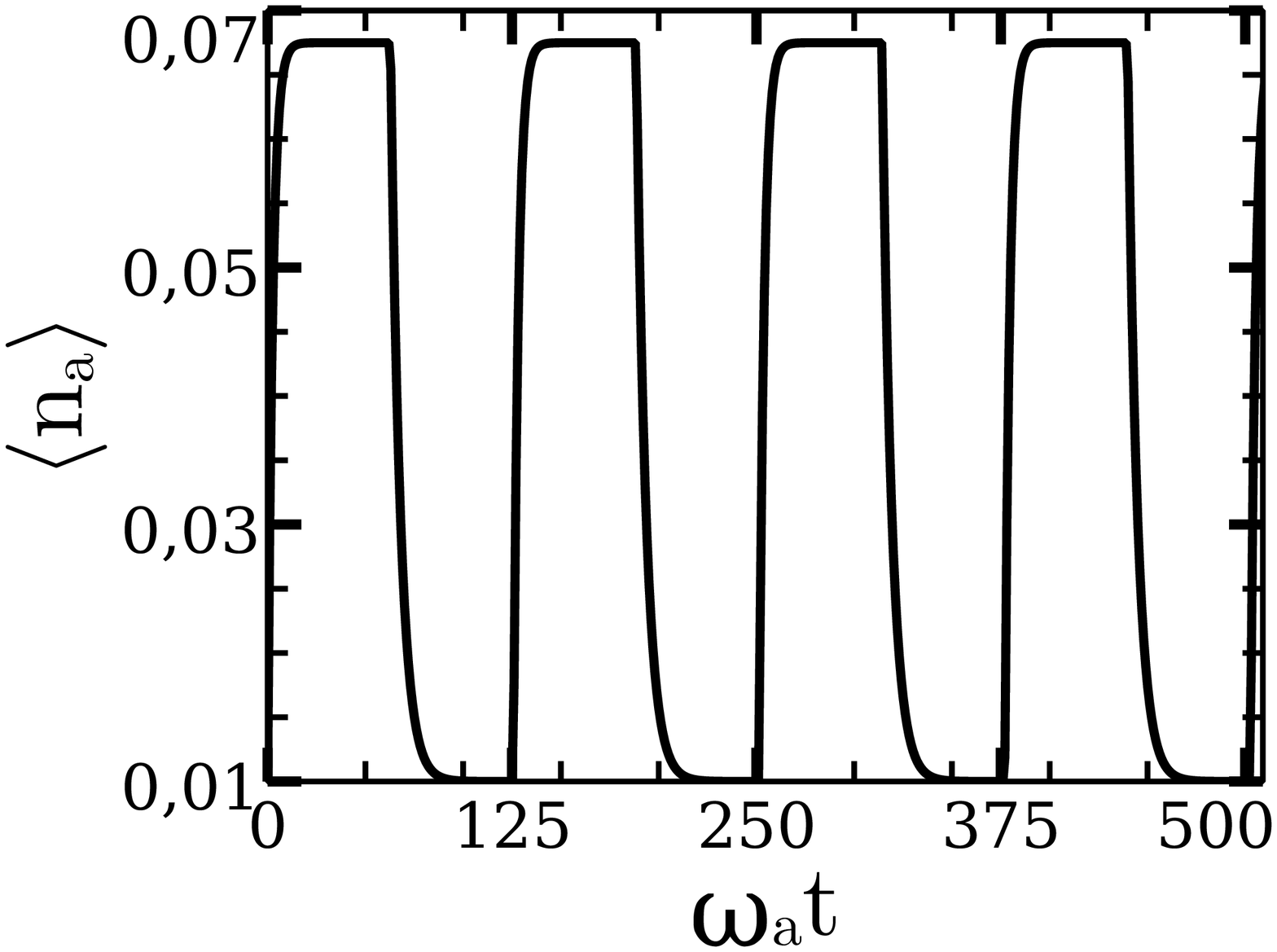}
		}\\
		\subfigure[]{
			\label{fig:fig2b}
			\includegraphics[width=7.5cm]{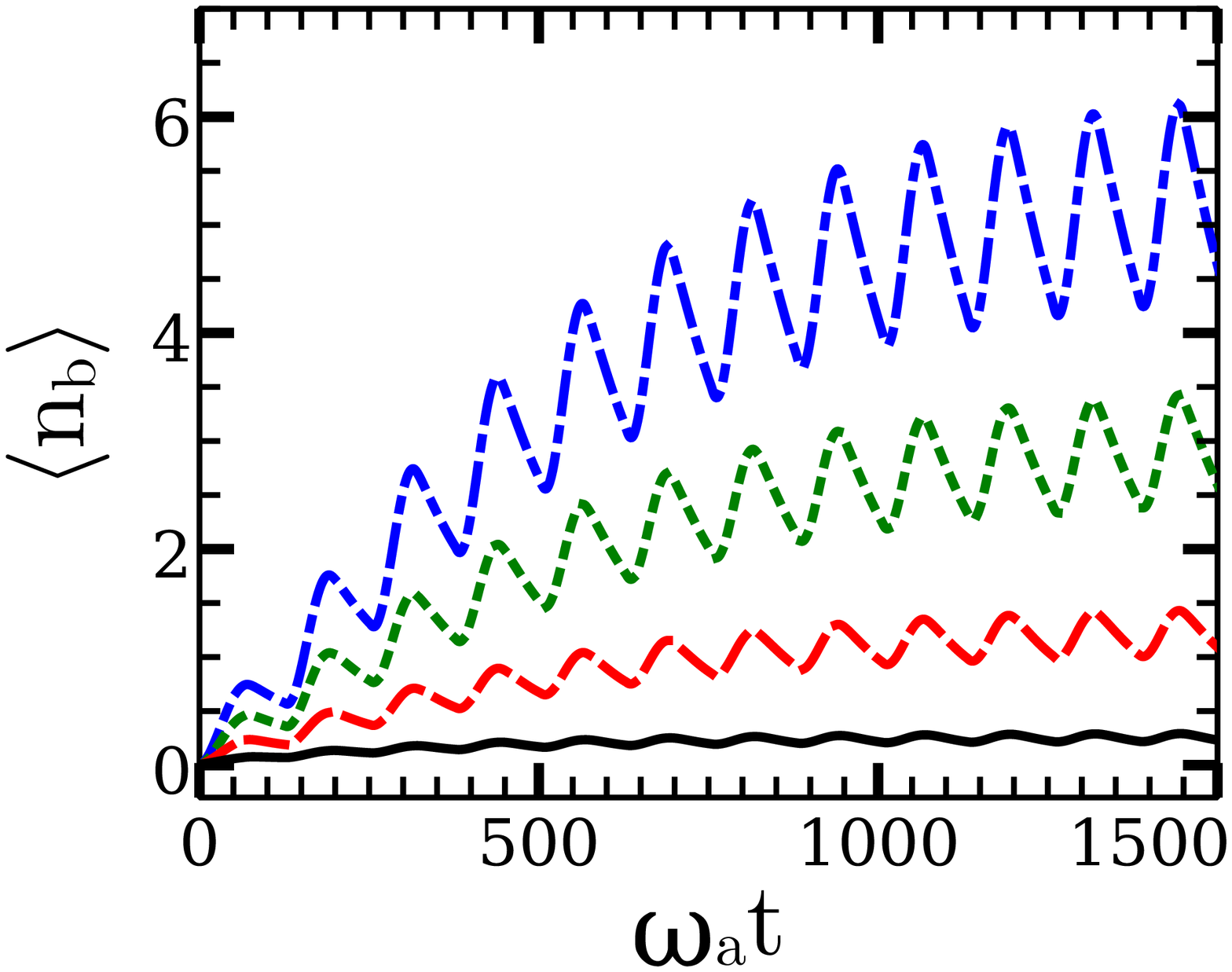}
		}\\
		\subfigure[]{
		\label{fig:fig2c}
		\includegraphics[width=7.5cm]{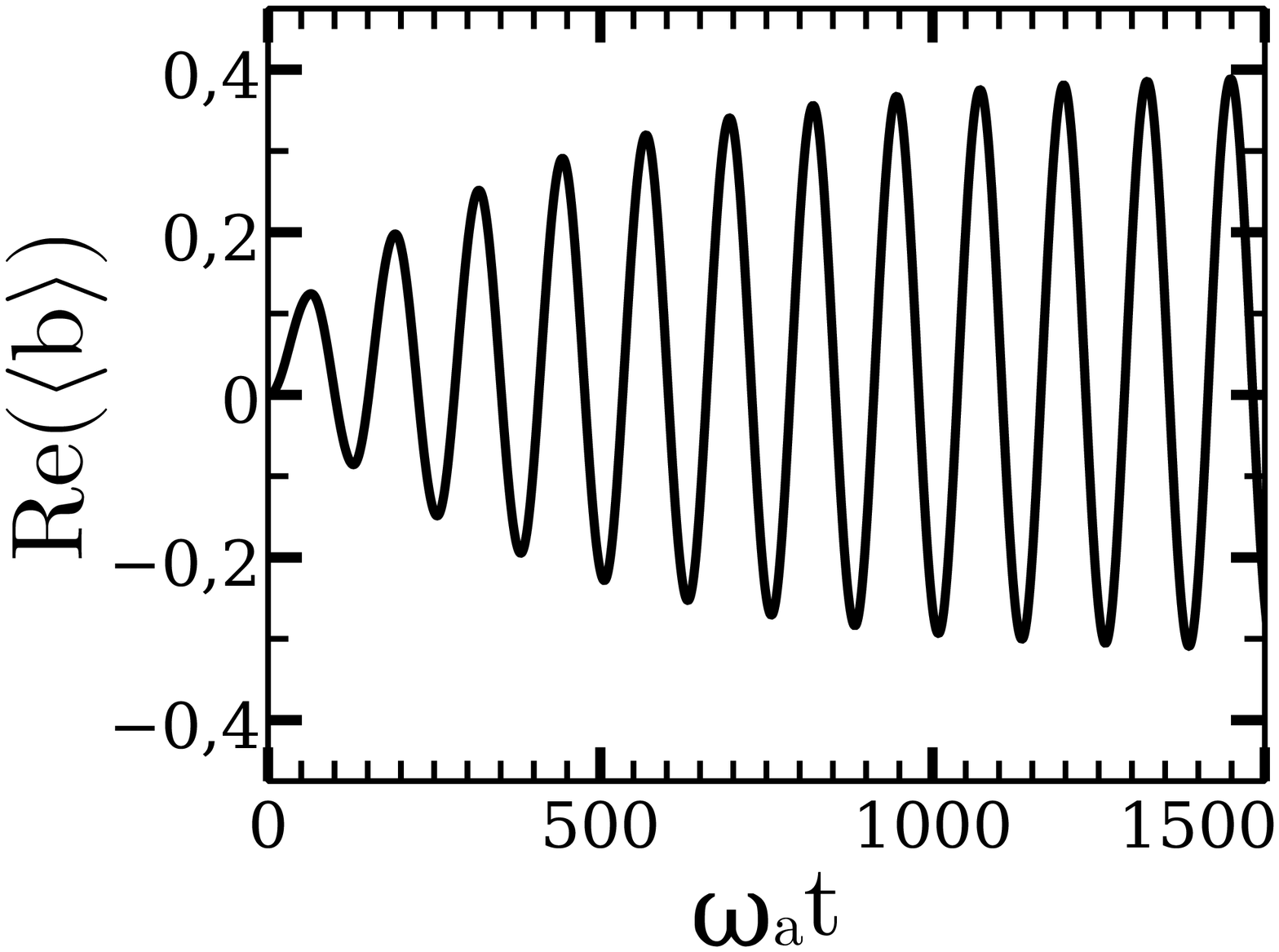}
	}
	\end{center}
	\caption{(Colour online) Dynamics of the mean number of excitations (a) $\langle \hat{n}_{a}\rangle$ and (b) $\langle \hat{n}_{b}\rangle$ 
	in the HF and LF modes, respectively, for $\bar{n}_{\text{h}}=0.125$ (black-solid), $\bar{n}_{\text{h}}=0.325$ (red-dashed),
	$\bar{n}_{\text{h}}=0.525$ (green-dotted) and $\bar{n}_{\text{h}}=0.725$ (blue-dot-dashed).(c) Dynamics of the LF field coherence is 
	characterized by $\Re(\langle b\rangle)$, where $\Re\equiv Re$ denotes the real part. 
	Time is   dimensionless and scaled with the HF mode frequency $\omega_a$.
	All the other parameters are as explained in the text.
	}
	\label{fig:fig2}
\end{figure}

In our simulations we use dimensionless parameters by scaling $\omega_{b}/2\pi=500$ MHz, $g/2\pi=500$ MHz, 
$\kappa_h/2\pi=\kappa_{a}/2\pi=2$ GHz and $\kappa_{b}/2\pi=50$ MHz by $\omega_{a}/2\pi=10$ GHz. 
The temporal profile of the incoherent drive acting on the HF mode is taken as a square wave $\kappa_{h}(t):=\kappa_{h}s(t)$.
The square wave $s(t)=1$ and $s(t)=0$ for the heating and cooling stages, respectively. Each stage takes the same time of $\pi/\omega_b$. In units of $1/\omega_a$ the cycle duration is then $t_{\text{cycle}}=(2\pi)20$ (cf. Fig.~\ref{fig:fig2}).

The mean number of excitations in the effective baths are taken to be $\bar n_a = \bar n_b := \bar{n}_{\text{c}}=0.01$ and $\bar n_h>0.1$, 
for which our simulations yield limit cycles for the
engine operation at steady state. The corresponding effective temperatures become $T_a \sim 104$ mK and
$T_b\sim 5$ mK. A typical environment temperature for the superconducting resonators is $T_0\sim 20$ mK. Accordingly
the hierarchy of temperatures associated with the engineered environments for the resonators becomes $T_h>T_a>T_0>T_b$. 
The temperature ranges of $T_h>T_a>T_0$ have been successfully produced experimentally 
for a single superconducting resonator using the noise drive method~\cite{fink_quantum-classical_2010}. The range of $T_0>T_b$
can be engineered using proposals for optomechanical schemes~\cite{teufel2011sideband}. The
HF resonator mode is
effectively heated to $T_h$ when the external periodic noise pulse is on, and when the pulse is off it effectively cools to $T_a$. The LF resonator is always coupled to an effective cold bath. 

The quantum dynamics of the coupled
resonators subject to such effective heating and cooling stages is investigated by solving Eq.~(\ref{eq:master}) using QuTiP~\cite{johansson2013qutip}. We neglect the last four terms associated with the background environment in Eq.~(\ref{eq:master}) 
by assuming $\kappa_a\gg \kappa_0^a$ and $\kappa_0^b\bar n_0^b>\kappa_b\bar n_b$.
After an intake stage, where the resonators are at their respective initial states, repeated action
of the the heating
and cooling stages will lead the system into a limit cycle that can be considered as a heat 
engine cycle with a net power output.

The system is assumed to be in an initial state $\hat\rho(0)=\hat\rho_a(0)\otimes\hat\rho_b(0)$, where 
\begin{eqnarray}
\hat\rho_a(0)&=&\frac{1}{\text{Tr}e^{-\omega_a\hat a^\dag \hat a/T_a}}e^{-\omega_a\hat a^\dag \hat a/T_a},\\
\hat\rho_b(0)&=&\frac{1}{\text{Tr}e^{-\omega_b\hat b^\dag \hat b/T_b}}e^{-\omega_b\hat b^\dag \hat b/T_b},
\end{eqnarray}
are the initial density matrix operators of the HF and LF modes, respectively. 
Initial occupations of the modes then are $\bar n_c$ since $\langle \hat n_x\rangle(0)=
\langle \hat x^\dag \hat x\rangle(0)=Tr(\hat\rho(0)\hat x^\dag \hat x)=\bar n_c=0.01$, with $x=a,b$.
\subsection{Quantum piston in phase space}
The heating and cooling stages in the engine operation of the coupled resonator system can be visualized by examining 
the dynamics of the mean excitation numbers 
$\langle \hat n_x\rangle =\langle \hat x^\dag \hat x\rangle=Tr(\hat\rho(t)\hat x^\dag \hat x)$, with $x=a,b$, 
which are shown in Fig.~\ref{fig:fig2}. The dynamics of $\langle \hat n_a\rangle$ is shown in Fig.~\ref{fig:fig2a} for $\bar{n}_{\text{h}}=0.125$.  
The heating pulse duration is taken to be longer than the thermalization time of the HF  resonator.
When the heating pulse is applied, the HF mode reaches a steady state
at $\langle \hat n_a\rangle_{\text{ss}}\sim 0.0675$, which is the same as the analytical result 
$\langle \hat n_a\rangle_{\text{ss}}=(\bar n_c+\bar{n}_{\text{h}})/2$ (see the Appendix~\ref{app:q-steadyState}).
After the heating pulse, the excitation number first rapidly drops and then 
slowly cools back to the initial value. Analytical expressions can be approximately given as
\begin{eqnarray}\label{eq:na_soln}
 \langle \hat n_a\rangle=\begin{cases}
 \frac{\bar n_a+\bar n_h}{2}-\frac{\bar n_h-\bar n_a}{2}e^{-(\kappa_a+\kappa_h)t}, & 0\le t<\frac{\pi}{\omega_b}; \\
 \bar n_a+\frac{\bar n_h-\bar n_a}{2}e^{-\kappa_a (t-\pi/\omega_b)}, & \frac{\pi}{\omega_b}\le t<\frac{2\pi}{\omega_b},
 \end{cases}
 \end{eqnarray}
 which is repeated indefinitely.
\begin{figure*}[!t]
\centering
\begin{center}
		\subfigure[]{
			\label{fig:fig3a}
			\includegraphics[width=4.15cm]{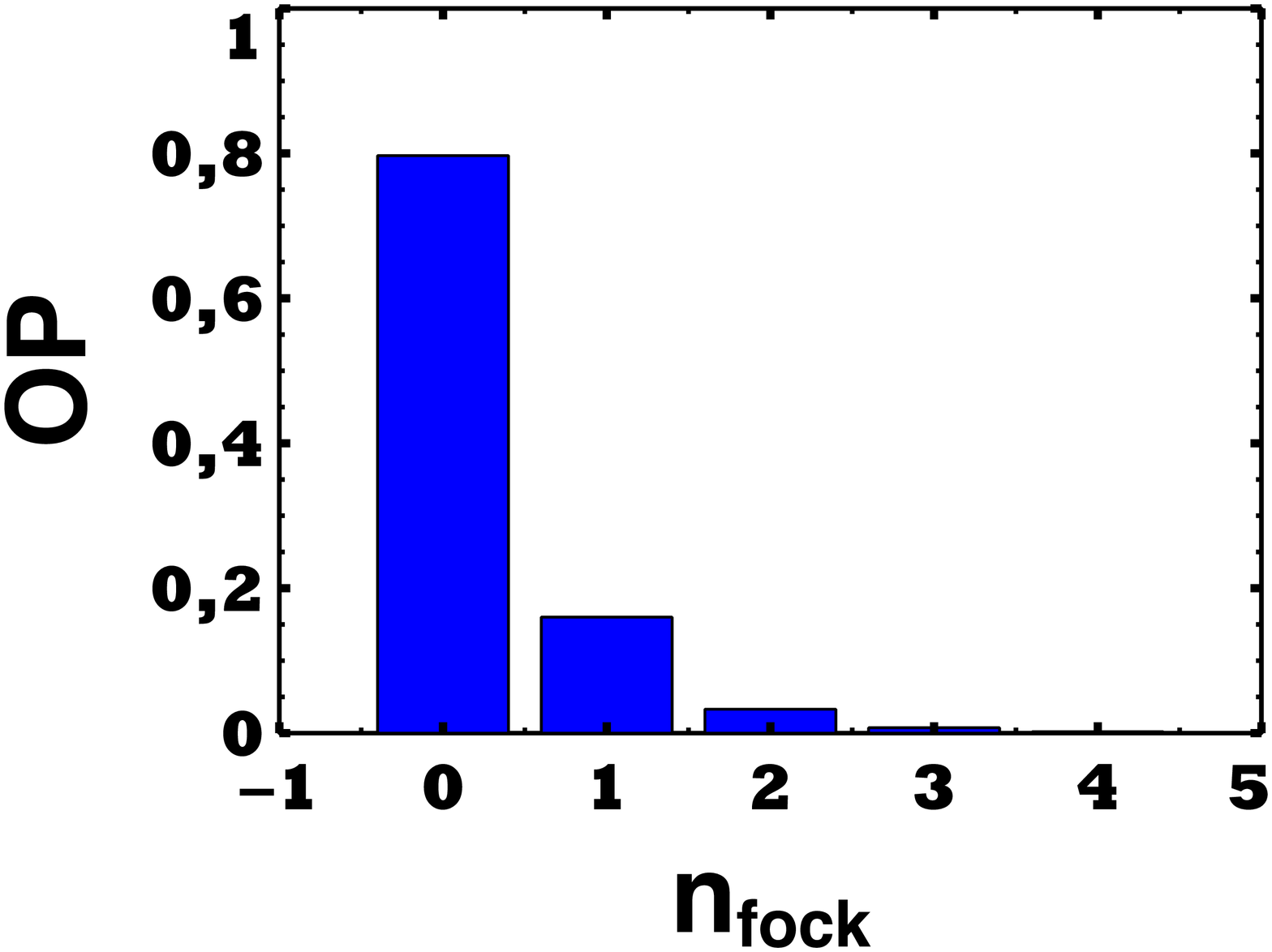}
		}
		\subfigure[]{
			\label{fig:fig3b}
			\includegraphics[width=4.15cm]{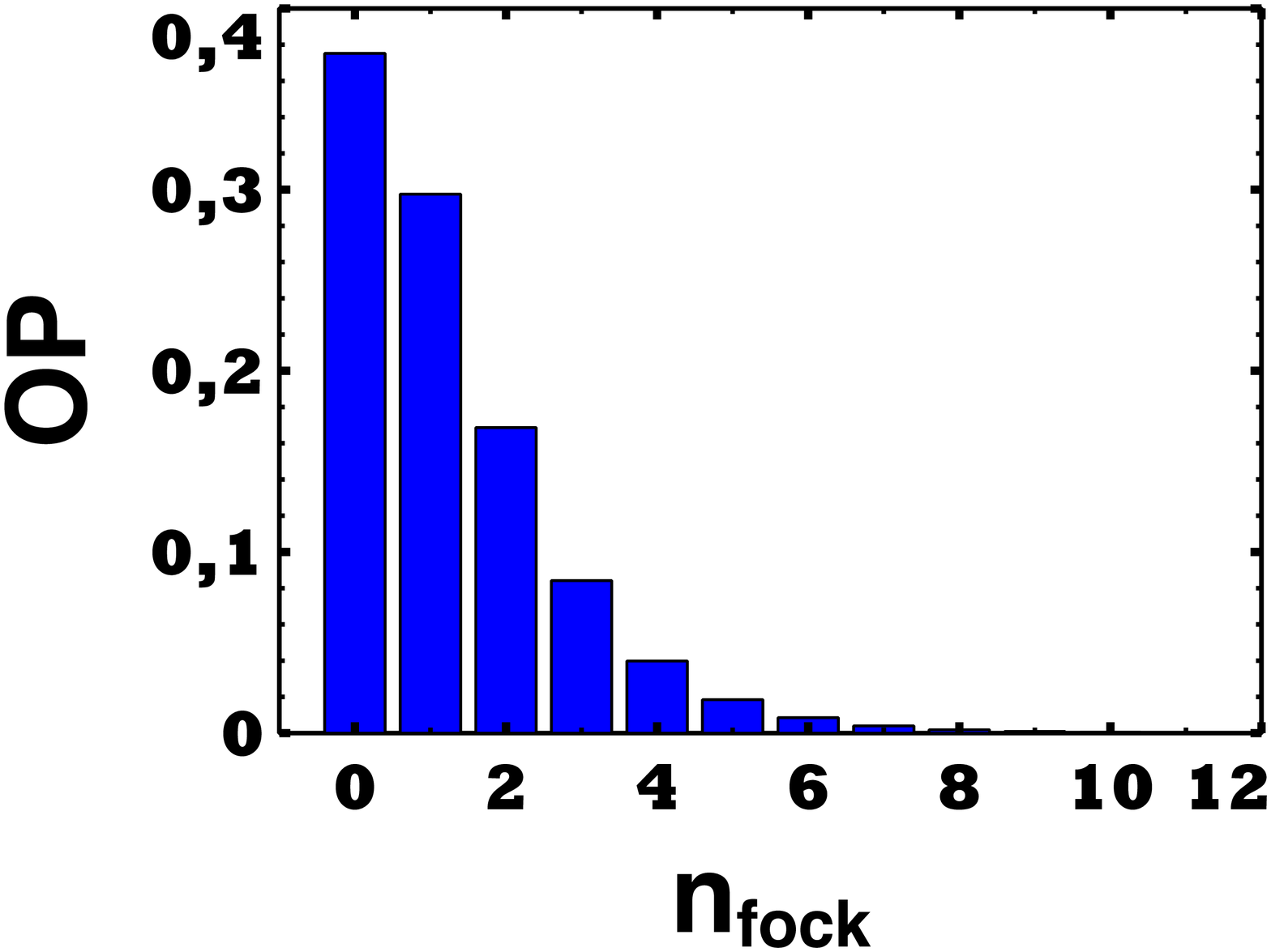}
		}
		\subfigure[]{
			\label{fig:fig3c}
			\includegraphics[width=4.15cm]{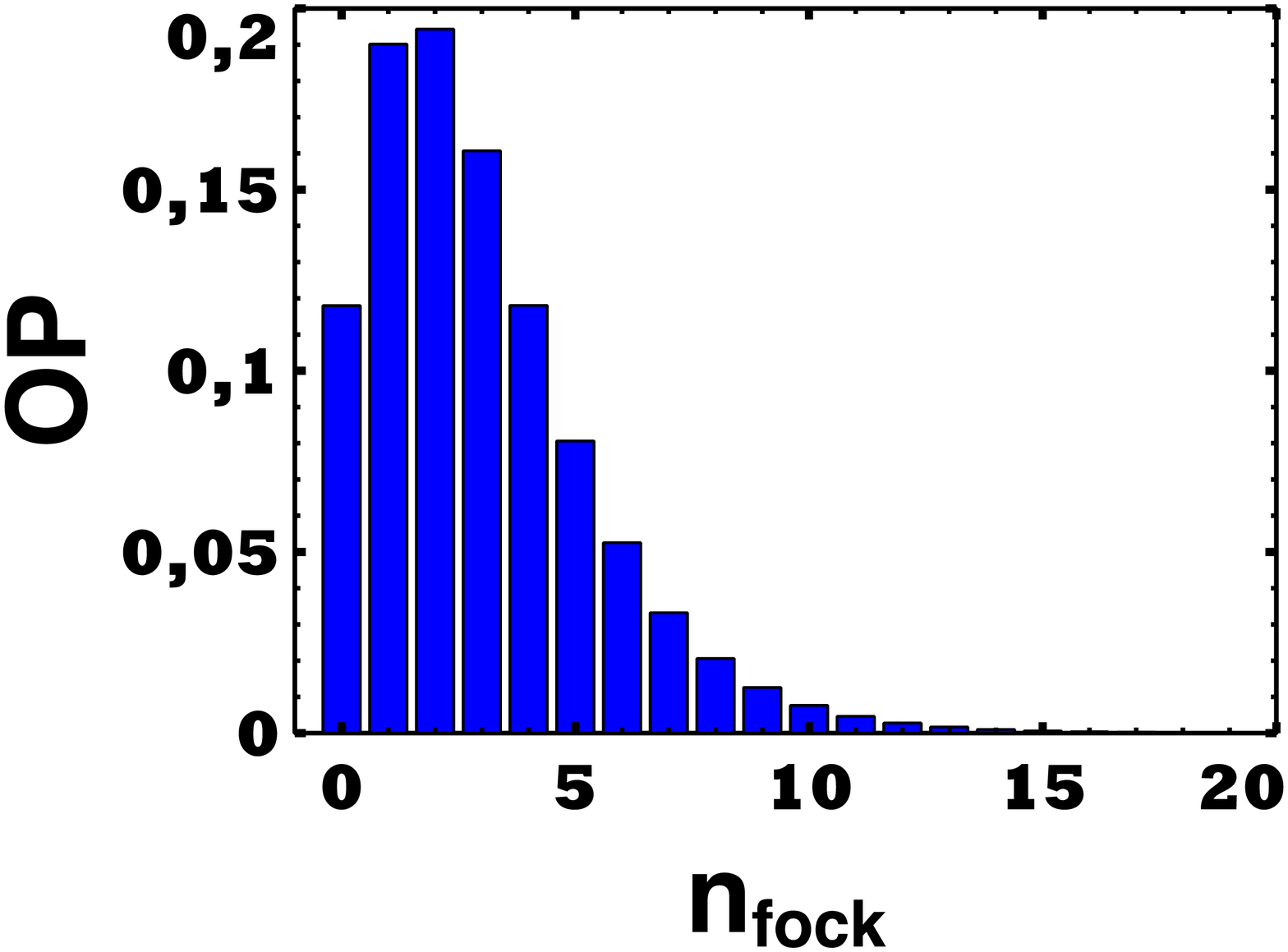}
		}
		\subfigure[]{
			\label{fig:fig3d}
			\includegraphics[width=4.15cm]{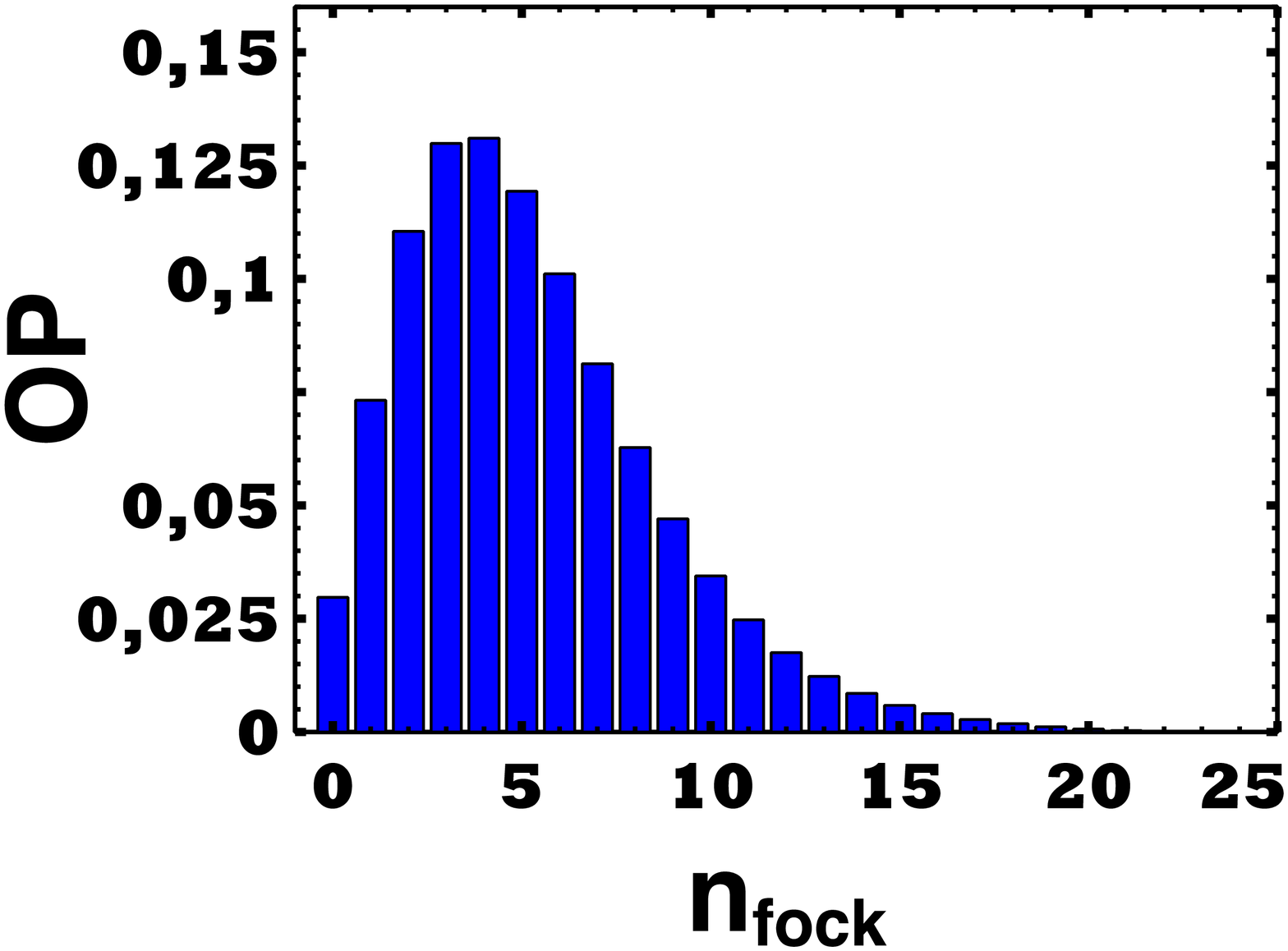}
		}
\end{center}
\caption{(Colour online) (a)-(d) The occupation probability (OP) with respect to occupation number 
of the Fock states $n_{\text{fock}}$ of the LF resonator mode at 
$\omega_{\text{a}}t=5000$
for $\bar{n}_{\text{h}}=0.125$ (a), $\bar{n}_{\text{h}}=0.325$ (b), 
$\bar{n}_{\text{h}}=0.525$ (c) and 
$\bar{n}_{\text{h}}=0.725$ (d). All the other parameters are as explained in the text.}
\label{fig:fig3}
\end{figure*}

The dynamics of the mean excitations of the LF mode $\langle\hat n_b\rangle$ are shown in Fig.~\ref{fig:fig2b}, 
which reaches a steady state with coherent oscillations in
the long time limit. Both the time average $\langle\hat n_b\rangle$ value and the amplitude of the oscillations are larger for larger $\bar{n}_{\text{h}}$; 
while the frequency of oscillations remains the same.
The heating pulse duration
is only long enough to thermalize the HF resonator, while the LF resonator is unable to equilibrate in a pulse period. The evolution of the
LF resonator will start from a different initial condition for every pulse action. 
This ``memory" effect leads to a higher time average $\langle\hat n_b\rangle$ value than the one found under continuous drive (cf. Appendix).
In addition, correlations between the HF and LF resonator modes contribute to 
to the dynamics of the LF mode excitations. We will explore the physical mechanisms behind the dynamics 
of $\langle n_b\rangle$ in more detail in Sec.~\ref{sec:classicalVSquantum}.

Fig.~\ref{fig:fig2c} plots the ``electrical displacement'' of the LF mode, $\Re(\langle\hat{b}\rangle)=q/2$, where 
$q=\langle \hat q\rangle=\langle\hat b+\hat b^\dag\rangle$,
showing coherent oscillations in the steady state. The notation $\Re(\cdot)$ stands for the real part. In contrast to the dynamics of 
$\langle n_b\rangle$, $q$ has identical classical and quantum dynamics. Its equation of motion (see Appendix) is that of a 
periodically driven damped oscillator, which is given by
\begin{eqnarray}\label{eq:oscillator}
\ddot{q}+\kappa_b\dot{q}+\omega^2q=2\omega_bg\langle n_a(t)\rangle,
\end{eqnarray}
where $\omega^2=\omega_b^2+\kappa_b^2/4$ and the dots over $q$ indicate time derivatives. This equation can be interpreted in terms of an
effective series RLC circuit. The cold environment of the LF resonator acts as the resistive element dampening the LC oscillations. In the standard way, the
natural frequency $\omega_b$ of the LC oscillations is further renormalized by the damping rate $\kappa_b$. The
HF resonator provides the input 
voltage which sustains the oscillations. Despite the thermal nature of the HF resonator, it drives the RLC circuit into coherent
oscillations because the drive only depends on $\langle n_a\rangle$,
which is alternating between high and low values periodically. We note that while $q$ is not vanishing in the steady state, the electrical momentum $p=\langle\hat p\rangle=i\langle(\hat b^\dag-\hat b)\rangle$
becomes approximately zero (It is exactly zero according to the global master equation as can be seen
in the Appendix~\ref{app:q-steadyState}). 

The LF oscillator is in the weak damping regime ($\kappa_b<\omega_b$). The formal solution of Eq.~(\ref{eq:oscillator}) is given by
\begin{eqnarray}\label{eq:solution}
q(t)=2g\int_{0}^{t}\text{d}\,t^{\prime}\langle n_a(t^{\prime})\rangle e^{-\frac{\kappa_b}{2}(t-t^{\prime})}\sin{(\omega_b(t-t^{\prime}))}.
\end{eqnarray}
The periodic nature of $\langle n_a(t^{\prime})\rangle$ leads to an intuitive understanding of the emergence of a coherent steady state
in $q(t)$. Harmonics of $\langle n_a(t^{\prime})\rangle$ are given by the frequencies $\omega_k=k\omega_b$ where $k$ is an integer. According 
to Eq.~(\ref{eq:solution}) the maximum overlap or the resonance would occur for the first harmonic $k=1$. Besides, the higher harmonics
would have relatively smaller significance as their amplitudes get smaller with $k$. Accordingly, steady state oscillations in $q(t)$ are dominated by
the single frequency $\omega_b$. The time average value of $q$, as well as the amplitude of oscillations, increase linearly with $\bar n_h$. In terms
of our physical parameters, the approximate analytical solution is found to be
\begin{eqnarray}
q=\frac{\bar n_h+3\bar n_c}{2 }+\frac{20 (\bar n_h-\bar n_c)}{\pi } \sin \left( \omega _b t-1.55\right).
\end{eqnarray}
A more general solution is given in the Appendix~\ref{app:q-GenSteadyState}.
\begin{figure}[!t]
            \includegraphics[width=7.5cm]{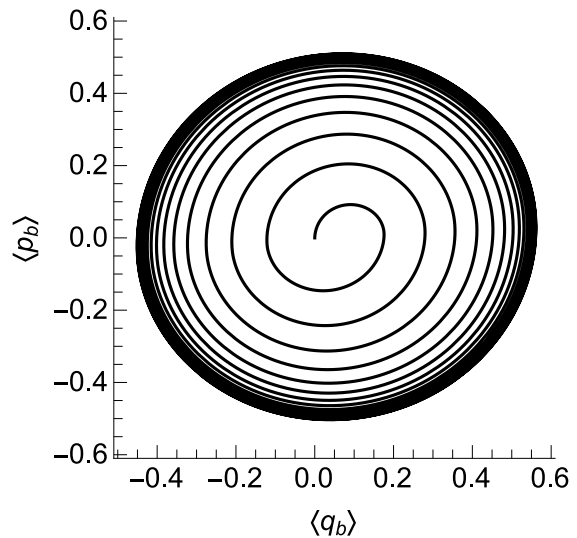}
                \caption{(Colour online) Phase portrait $\langle q_{b}\rangle-\langle p_{b}\rangle$ showing the dynamics of the LF mode 
                for $\bar{n}_{\text{h}}=0.125$. All the other parameters are as explained in the text.
                }
   \label{fig:fig4}
\end{figure}

We further visualize the coherence in the steady state of the LF mode by investigating the probability distribution 
$P(n):=\langle n|\rho_b(t)|n\rangle$, where $\rho_b(t) = \text{Tr}_a[\rho(t)]$ is the reduced density matrix of the LF mode, in Fig.~\ref{fig:fig3}. As we increase $\bar{n}_{\text{h}}$ from $\bar{n}_{\text{h}}=0.125$
to  $\bar{n}_{\text{h}}=0.725$, the the probability distribution 
changes from a thermal distribution to a coherent distribution, as shown in Figs.~\ref{fig:fig3a}-\ref{fig:fig3d}.

In Fig.~\ref{fig:fig4}, we plot the dynamics of the mean values of the field quadratures 
$q_b:=\langle \hat{q}_{\textbf{b}}\rangle:=q/\sqrt{2}$ and 
$p_b:=\langle \hat{p}_{\textbf{b}}\rangle:=p/\sqrt{2}$, where $p=\langle\hat p\rangle:=i\langle(\hat{b}^{\dagger}-\hat{b})\rangle$, with respect 
to each other. The resulting phase diagram is that of a periodically 
driven damped displaced harmonic oscillator, where the slight shift of the center
of the limit cycle from the origin is the signature of the coherent displacement induced by the thermal drive.
The HF mode reaches steady state within a single thermal noise pulse duration as shown in Fig.~\ref{fig:fig2a}. The action of the thermal 
noise pulse on the HF mode is translated to the LF mode by the optomechanical coupling. According to 
Fig.~\ref{fig:fig2b}, the LF mode can reach an oscillatory steady state after the action of a several noise pulses.
As the LF mode lags behind the HF mode, there will be a transient regime before a limit cycle 
is established, as we see in In Fig.~\ref{fig:fig4}. 
While a limit cycle emerges for our model, we note that the existence of a stable 
limit cycle after transients is in general model and initial condition dependent~\cite{feldmann_characteristics_2004}. Within the approximation of
$\langle\hat n_a\rangle$ as a square wave drive, we find an 
expression for the limit cycle (see Appendix~\ref{app:q-GenSteadyState})
\begin{eqnarray}
(p_b-p_{b0})^2+(q_b-q_{b0})^2=R^2,
\end{eqnarray}
where
\begin{eqnarray}
R&=&\frac{4}{\pi}\frac{\sqrt{2}\omega_bg}{\sqrt{\kappa_b^2\left(\omega_b^2+\frac{\kappa_b^2}{16}\right)}}
\left(\frac{\bar n_h-\bar n_c}{4}\right),\\
q_{b0}&=&\frac{2\omega_bg}{\sqrt{2}\omega^2}\left(\frac{\bar n_h+3\bar n_c}{4}\right),
\end{eqnarray}
and $p_{b0}\approx 0$.
\begin{figure}[!t]
            \includegraphics[width=7.5cm]{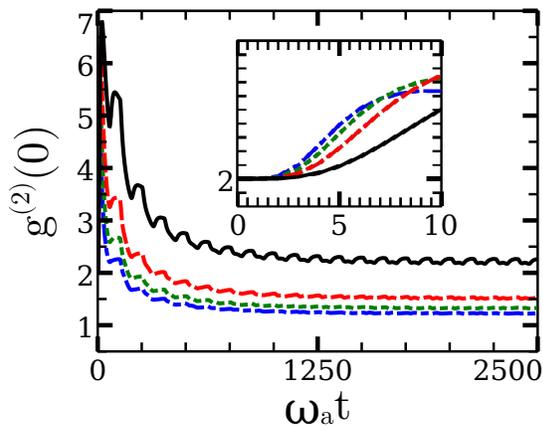}
                \caption{(Colour online) Time dependence of the zero-time delay second order correlation function $g^{(2)}(0)$ of the 
                LF mode for $\bar{n}_{\text{h}}=0.125$ (black-solid), $\bar{n}_{\text{h}}=0.325$ (red-dashed), $\bar{n}_{\text{h}}=0.525$
                 (green-dotted) and $\bar{n}_{\text{h}}=0.725$ (blue-dot-dashed). The inset is the magnification of $g^{(2)}(0)$ 
                 up to $10\omega_{a}t$. Time is   dimensionless and scaled with the HF mode frequency $\omega_a$. 
                 All the other parameters are as explained in the text.
                }
   \label{fig:fig5}
\end{figure}

The effect of coherence in the LF mode state can be further revealed by calculating the zero-time delay second order correlation function 
$g^{(2)}(0):=\langle\hat{b}^{\dagger}\hat{b}^{\dagger}\hat{b}\hat{b}\rangle/\langle \hat{n}_{b}\rangle^2$, of the LF mode.
Fig.~\ref{fig:fig5} shows the time dependence of $g^{(2)}(0)$ of the LF mode for the same $\bar{n}_{\text{h}}$ 
values as in Fig.~\ref{fig:fig2}. Initially, $g^{(2)}(0)=2$ for all cases as the LF mode starts in a thermal state. $g^{(2)}(0)$ decrease as the LF mode gains partial coherence in time. At higher $\bar{n}_{\text{h}}$, the steady state value of $g^{(2)}(0)$ gets closer to
the coherent state value $g^{(2)}(0)=1$. While perfect coherence is not achieved within the range of values of $\bar{n}_{\text{h}}$ considered, a
slow convergence to coherent state statistics can be seen in Fig.~\ref{fig:fig5}. We note that 
the second-order coherence function can
be measured in circuit QED systems using various techniques such as by linear detectors~\cite{da_silva_schemes_2010}.
\subsection{Semi-classical Engine cycle}
\begin{figure}[!t]
	\centering
	\begin{center}
		\subfigure[]{
			\label{fig:fig6a}
			\includegraphics[width=7.5cm]{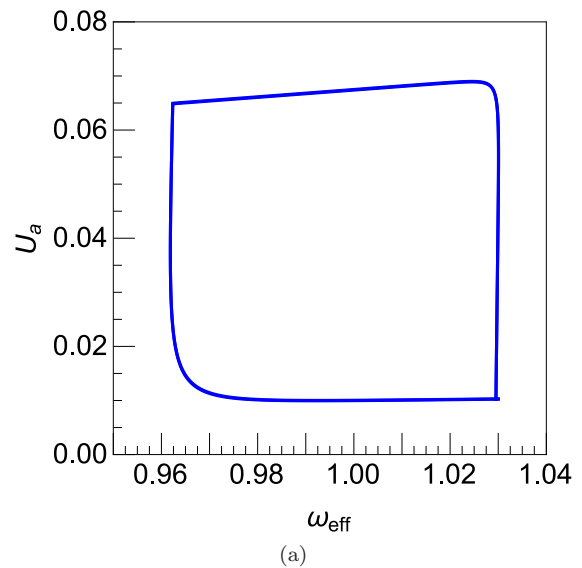}
		}\\
		\subfigure[]{
			\label{fig:fig6b}
			\includegraphics[width=7.5cm]{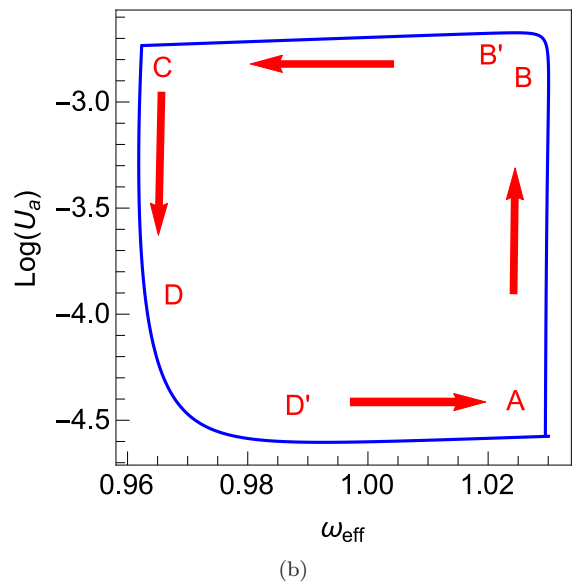}
		}
	\end{center}
\caption{(Colour online) (a) Dependence of the effective internal energy of the HF resonator 
$U_a=\omega_{\text{eff}}\langle\hat n_a\rangle$ on the effective HF mode frequency
$\omega_{\text{eff}}=\omega_a-g(\langle \hat b\rangle+\langle \hat b^\dag\rangle)$ for 
$\bar{n}_{\text{h}}=0.125$ in the steady state. (b) The plot shows $\ln{(U_a)}$ in order to 
emphasize that the temperature in the lower branch (D$^{\prime}$ to A) is increasing. 
The direction of the engine cycle in the
steady state is indicated by the red arrows. 
The points A, B, B$^{\prime}$, C, D, and D$^{\prime}$ are used in the text 
to describe the thermodynamic processes
in the cycle. We use the frequency
of the HF resonator, $\omega_a$, to scale our parameters. 
All the other parameters are as explained in the text.}
\label{fig:fig6}
\end{figure}
In order to describe the engine cycle, we introduce an effective HF mode 
frequency $\omega_{\text{eff}}:=\omega_a-gq$,
which can be interpreted as the change in the frequency associated with the variations in the electrical length of the HF resonator. Accordingly, the effective
mean energy of the working fluid can be taken to be $U_a=\omega_{\text{eff}}\langle\hat n_a\rangle$. This semiclassical factorization ignores the
correlations between $\hat n_a$ and $\hat q$. When we calculate $U_a$ without the factorization assumption, 
we find qualitatively the same cycle
pictures and the work output is negligibly enhanced. For the moment, we consider the 
semiclassical explanation of the engine cycle and
discuss the effect of quantum correlations separately in the later sections.

\begin{figure}[!t]
 \includegraphics[width=7.5cm]{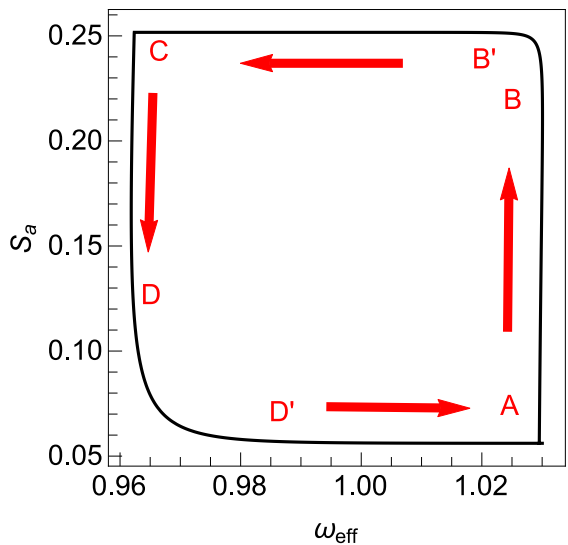}
  \caption{(Colour online) Steady state dependence of the entropy $S(\rho_{a})$ on 
the effective HF mode frequency 
  $\omega_{\text{eff}}=\omega_a-g(\langle \hat b\rangle+\langle \hat b^\dag\rangle)$ for $\bar{n}_{\text{h}}=0.125$. 
  We use the frequency
    of the HF resonator $\omega_a$ to scale our parameters.
    All the other parameters are as explained in the text.
                }
   \label{fig:fig7}
\end{figure}
In Fig.~\ref{fig:fig6}, we plot the dependence of $U_a$ on $\omega_{\text{eff}}$ in the steady state at $\bar{n}_{\text{h}}=0.125$,  for 
the same set of parameters used in earlier figures. We plot $\ln{(U_a)}$ in order to 
emphasize that the temperature in the lower branch (D$^{\prime}$ to A) is increasing. 
A four stage engine cycle can be identified in this picture.

The first stage is indicated by the arrow from point A to B in Fig.~\ref{fig:fig6} at $\omega_{\text{eff}}\sim 1.03$ and corresponds
to an isochoric heating of the HF resonator by the incoming heat pulse. 
The electrical length $1/\omega_{\text{eff}}$ remains constant while the incoherent energy
is received from the external noise pulse. 
The coherence of the LF mode cannot follow the thermalization of the HF mode as fast (cf. Fig.~\ref{fig:fig2}) so that the ``piston'', or the 
LF mode quadrature, remains at rest in the phase space. There is a transitional stage from B to B$^{\prime}$ which cannot be identified
with the standard thermodynamical processes. 
\begin{figure}[!t]
\includegraphics[width=7.5cm]{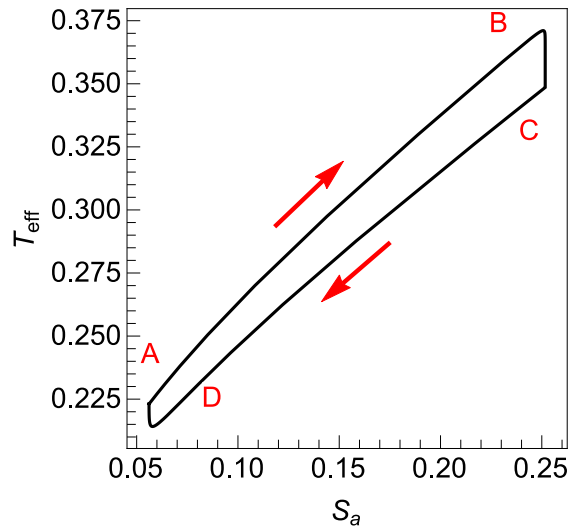}
\caption{(Colour online) $T-S$ diagram of the 
HF mode for $\bar{n}_{\text{h}}=0.125$. 
We use the frequency
of the HF resonator, $\omega_a$, to scale our parameters.
All the other  parameters are as explained in the text.}
\label{fig:fig8}
\end{figure}

The second stage, from B$^{\prime}$ to C, is adiabatic expansion of the HF resonator, 
where $\omega_{\text{eff}}$ decreases to $\sim 0.96$ and hence the electrical length of the HF resonator increases.
The entropy of the HF resonator field can be calculated by
\begin{eqnarray}
S_a=(1+\langle \hat n_a\rangle)\ln{(1+\langle \hat n_a)}-\langle \hat n_a\rangle\ln{(\langle \hat n_a)},
\end{eqnarray}
which remains constant in this stage as shown in Fig.~\ref{fig:fig7}. The heating pulse is still active, but the HF
mode is in thermal equilibrium and working on the LF mode. The coherence in the LF mode builds up in this stage as
the ``piston'' shifts in the phase space, converting heat to potential energy to be harvested.

The third stage is isochoric cooling that happens from C to D at $\omega_{\text{eff}}\sim 0.96$, following the rapid decrease of the population of the
HF mode when the heating pulse is turned off (cf. Fig.~\ref{fig:fig2}). There is another transitional stage from D to D$^{\prime}$. The final stage from D$^{\prime}$ to A closes the cycle. It corresponds to an adiabatic compression (cf. Fig.~\ref{fig:fig7}) where $\omega_{\text{eff}}$ increases to
$\sim 1.03$. The sign change in the coherence of the LF mode
(cf.~Fig.~\ref{fig:fig2c}) leads to an increasing $\omega_\text{eff}$ (decreasing electrical length) so that the ``piston'' moves back to its 
original location in the phase space and the cycle is complete.

\subsection{Performance of the engine}
Our 4-stage engine description, other than the transitional stages, can be considered as an
Otto cycle. The effects of the transitional stages do not strongly
influence the temperature-entropy (T-S) diagram plotted in Fig.~\ref{fig:fig8},
which closely resembles that of an Otto engine. Here we introduced an effective temperature given by
\begin{eqnarray}
T_{\text{eff}}=\omega_{\text{eff}}/\ln{(1+1/\langle\hat n_a\rangle)}.
\end{eqnarray}
The T-S diagram in Fig.~\ref{fig:fig8}, which follows a narrow cycle, is of a similar form to that obtained 
experimentally for the single-atom heat engine~\cite{rosnagel_single-atom_2016} and for a nanomechanical Otto
engine driven by a squeezed reservoir to operate with an efficiency beyond the classical Carnot limit~\cite{klaers_squeezed_2017}.  

The area of such diagrams can be considered as the potential work output from the working fluid. By approximating  the diagram in
Fig.~\ref{fig:fig8} by a trapezoid, we can estimate the net work by $W_a\sim3.5\times 10^{-3}\hbar\omega_a\sim 2.3\times 10^{-26}$ joules. 
Dividing by the cycle time, which is the heating pulse period $2\pi/\omega_b=2$ ns, we find the power from the working fluid as 
$P_a\sim 1.15\times10^{-17}$ Watts. The heat intake of the HF resonator can also be determined from the diagram and we find
$Q_{\text{in}}\sim 0.06\hbar\omega_a$ which yields an efficiency of 
$\eta=W_a/Q_{\text{in}}\sim 0.06\%$. Similar values can be consistently found from the cycle in $\langle\hat n_a\rangle$-$\omega_{\text{eff}}$
diagram, which is similar to Fig.~\ref{fig:fig6}. Approximating the
cycle by a rectangular path, we can verify that $W_a\sim\hbar (n_H-n_L)(\omega_H-\omega_L)$, where $n_H\sim 0.0675$ and $n_L=0.01$ (cf. Fig.~\ref{fig:fig2})
are the maximum and minimum $\langle \hat n_a\rangle$, respectively. Similarly, $\omega_H\sim 1.04$ and $\omega_L\sim 0.96$ (see e.g. Fig.~\ref{fig:fig7}) 
are the  maximum and minimum of $\omega_{\text{eff}}$, respectively. The heat intake in this case can be written as $Q_{\text{in}}\sim 
\hbar\omega_H(n_H-n_L)$.
Hence, the efficiency becomes that of an Otto engine $\eta=1-\omega_L/\omega_H\sim 0.08\%$ These values increase with $\bar n_h$. 
For example, at $\bar n_h=1$, we get $\eta\sim 0.27\%$, $P\sim 2.64\times 10^{-16}$ J/s, and $W\sim 5.28\times 10^{-25}$ J. We cannot
increase $\bar n_h$ indefinitely however. Optomechanical model requires relatively small oscillations such that $\omega_a>g|q|$
at all times. This limits our considerations to a regime $\bar n_h<3.25$.

\begin{figure}[!t]
\includegraphics[width=8.3cm]{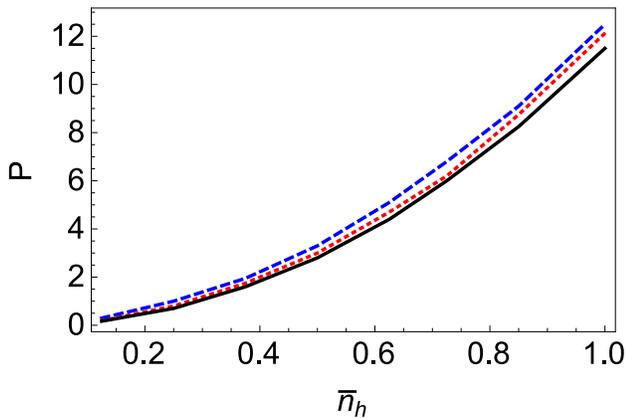}
\caption{(Colour online) The maximum power (in units of $\kappa_b\hbar\omega_a$)
with respect to average number of excitations $\bar{n}_{\text{h}}$ in the effective hot bath. 
The black solid, red dotted, and blue dashed curves indicate the results of the semiclassical, classical, and quantum calculations{\color{red}.} All the other parameters are as explained in the text.
	}
\label{fig:fig9}
\end{figure}
\begin{figure}[!t]
	\includegraphics[width=8.3cm]{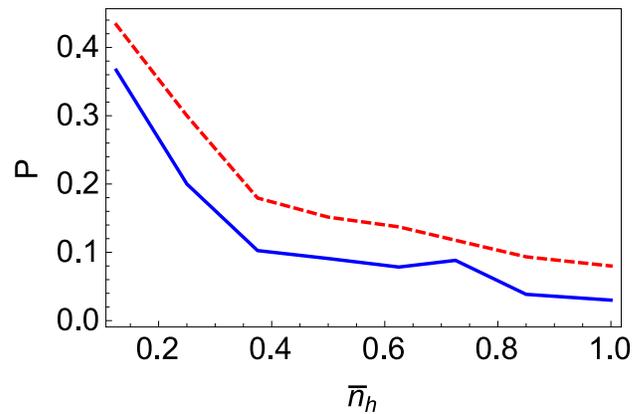}
	\caption{(Colour online) Behavior of the relative difference in the dissipative power with respect to average number of 
	excitations $\bar n_H$ in the hot bath. Red dashed and blue solid curves  indicate $(P_q-P_c)/P_q$ and $(P_q-P_{sc})/P_q$ where 
	$P_q,P_c$, and $P_{sc}$ are calculated by quantum, 
	classical, and semiclassical methods, respectively.
	Power is dimensionless and scaled with $\kappa_b\hbar\omega_a$. 
	All the other parameters are as explained in the text.}
	\label{fig:fig10}
\end{figure}

For practical purposes, a figure of merit called the dissipative power is also considered to 
estimate the performance of such piston engines, which is given by~\cite{mari_quantum_2015}
\begin{eqnarray}\label{eq:power}
P&=&-{\text{Tr}}\{\omega_b\hat n_b\kappa_{b}[(\bar{n}_{b}+1)D[\hat{b}]+\bar{n}_{b}D[\hat{b}^{\dagger}]]\}\\
&=& \omega_{b}\kappa_{b}\big(\langle\hat{n}_{b}\rangle-\bar{n}_{\text{c}}\big).
\end{eqnarray}
It is non-zero when the LF resonator mode 
is not in equilibrium with the environment. 
It has the same qualitative behaviour  with the mean number of excitations in the LF mode 
by definition and hence is oscillatory around a mean value (cf. Fig.~\ref{fig:fig2}). In steady state it has
a highly intuitive relation to the power output of the Otto cycle of the HF resonator. In order to see this
we note that $\langle \hat n_a\rangle=0$ in the adiabatic branches of the cycle so that the mean energy
$U_a=\omega_a\langle\hat n_a\rangle-g\langle\hat n_a\hat q\rangle$ varies as 
\begin{eqnarray}\label{eq:mechPower}
\frac{dU_a}{dt}=-g\omega_b\langle\hat n_a\hat p\rangle,
\end{eqnarray}
which is identical with $-P$ in the steady state (cf.~Appendix~\ref{app:q-steadyState}).
Here, we did not employ the factorization assumption and have used the unitary part of the 
equation of motion for the $\langle\hat n_a\hat q\rangle$, which can be
obtained from the equations of motion in Appendix~\ref{app:q-dynamics}. As the mean energy is 
determined by the usual force (pressure) times distance type expression, 
the associated power is also expressed by the typical force times velocity type formula. The
difference from the classical mechanics expressions is that quantum or classical statistical correlations
can contribute to the expectation values. $P$ is called the dissipative power because it is approximately the same as 
the heat current from the LF resonator to its thermal bath, which is given by
\begin{eqnarray}
J_b={\text{Tr}}[\hat{{\cal L}}_b\hat H_{\text{sys}}]
=P+g\frac{\kappa_b}{2}\langle\hat n_a\hat q\rangle,
\end{eqnarray}
where $\hat{{\cal L}}_b$ represents all the Lindblad superoperators depending on $\kappa_b$. 
The last term in the $J_b$ expression is negligible. The approximation of $J_b\approx P$ 
improves by including
the non-local effects of the reservoir-system interactions. The last two statements are verified
numerically for our parameter regimes using the global master equation given in Appendix~\ref{app:q-dynamics}. 

The thermal current measurement into the reservoir
of the LF resonator then can be used to determine the mechanical power output of the Otto
cycle of the system. Alternatively, the mechanical output could be harvested 
directly by attaching an electrical load to the open terminal of the LF transmission line
resonator~\cite{mari_quantum_2015}. The mechanical power of the HF resonator is used 
to ``charge" the LF resonator
by putting it into a thermal coherent state. Such states belong to the class of so called 
thermodynamically non-passive states
and are capable of producing useful work. Uncoupling the LF resonator from the system
would allow it to be used as a quantum coherent resource~\cite{mari_quantum_2015,gelbwaser-klimovsky_work_2015}. As 
thermal states are passive states,
the dissipative work can also be used to characterize the non-passivity of the LF resonator.

We report the maximum power $P_{\text{max}}$, which is 
evaluated for the maximum $\langle n_b\rangle$ in the steady state, in Fig.~\ref{fig:fig9}. After $\bar n_h\sim 0.325$ the coherent character of the 
LF mode steady state becomes more significant than its thermal character (cf.~Fig.~\ref{fig:fig5}), and hence the dissipated power has more useful 
work content than the incoherent energy. The other two curves in the figure corresponds to the results of the semiclassical and classical
models of the system. The semiclassical model, as described in the preceding subsection about the engine cycle, ignores the quantum correlations
and factorizes the two operator products $\langle \hat n_a q\rangle$ and $\langle \hat n_a p\rangle$ (see their equations of motion in the Appendix.). 
We
will describe the classical model next and explain the hierarchy of the curves in Fig.~\ref{fig:fig9} in terms of the classical and quantum correlations. We note that as $\bar n_h$ grows, the quantum correlations are less significant. 
The relative difference between the classical and
quantum descriptions diminish with increasing temperature as shown in Fig.~\ref{fig:fig10}.
\begin{figure}[!t]
	\includegraphics[width=7.5cm]{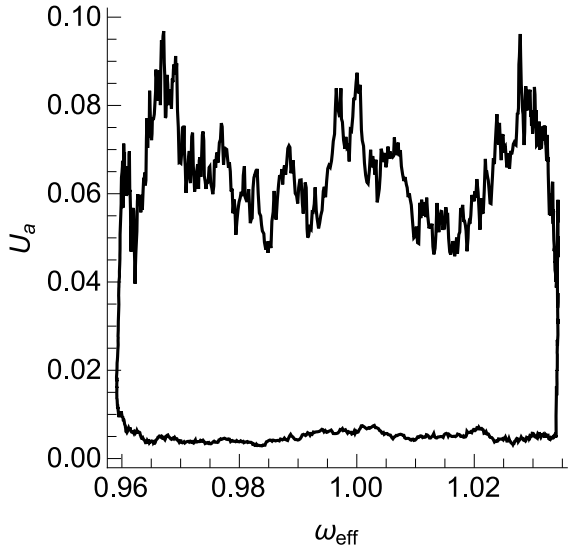}
	\caption{(Colour online) Change in the internal energy $U$ of the classical engine with respect to the effective frequency $\omega_{\text{eff}}$ 
	for $\bar{n}_{\text{h}}=0.125$. We use the frequency
    of the HF resonator $\omega_a$ to make our parameters scaled and dimensionless. 
    All the other parameters are as explained in the text.
	}
	\label{fig:fig11}
\end{figure}
\section{Classical model}\label{sec:classicalModel}
In order to distinguish quantum features in our engine system from its classical counterpart, we shall now treat the model Hamiltonian
in Eq.~\ref{eq:model} as a classical model. Replacing the operators of the fields of the resonators with 
the $c$-numbers such that $\hat{a}\rightarrow\alpha_{a}$ and $\hat{b}\rightarrow\alpha_{b}$ we get the classical Langevin equations
\begin{eqnarray}
\dot{\alpha}_{a}&=&-\left(i\omega_{a}-ig(\alpha_{b}^{*}+\alpha_{b})+\frac{\kappa_{h}}{2}+\frac{\kappa_{a}}{2}\right)\alpha_{a} \nonumber\\
&+&\xi_{h}(t)+\xi_{a}(t),\\
\dot{\alpha}_{b}&=&-\left(i\omega_{b}+\frac{\kappa_{b}}{2}\right)\alpha_{b}+ig|\alpha_{a}|^2+\xi_{b}(t),
\end{eqnarray}
where 
$\xi_i(t)$ with $i=a,b,h$ represents time dependent delta-correlated stochastic noise with $\langle\xi_i(t)\rangle=0$ and $\langle\xi_i(t_1)\xi_i(t_2)\rangle:=D_i\delta(t_1-t_2)$ where $D_i$ is the strength of the noise. The parameter $D_i$ is a 
function of $\kappa_i$ according to the fluctuation-dissipation theorem~\cite{kubo1966fluctuation} such that $D_i=\kappa_i \bar{n}_i$.
\begin{figure}[!t]
	\includegraphics[width=7.5cm]{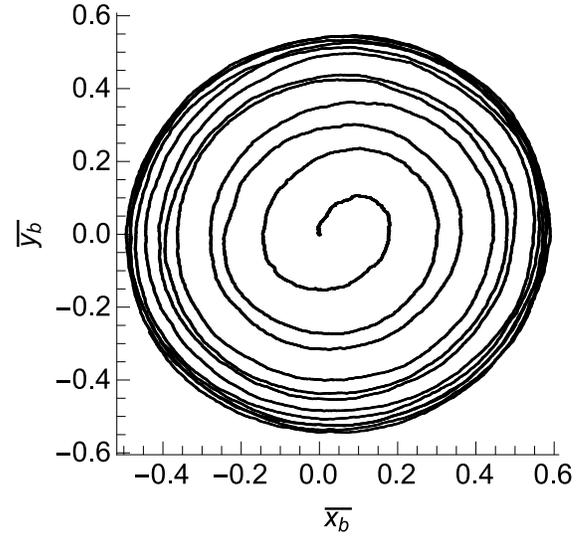}
	\caption{(Colour online) Phase portrait $X_{b}-Y_{b}$ of the classical LM resonator for $\bar{n}_{\text{h}}=0.125$. All the other parameters 
	are as explained in the text.
	}
	\label{fig:fig12}
\end{figure}

We define the field quadratures $X_i,Y_i$ such that $\alpha_{a}:=1/\sqrt{2}(X_{a}+iY_{a})$ and $\alpha_{b}:=1/\sqrt{2}(X_{b}+iY_{b})$. 
By writing the noise parameters as $\xi_i=1/\sqrt{2}(\xi_{x}^{i}+i\xi_{y}^{i})$, we can express the equations to be simulated as
\begin{eqnarray}
\nonumber dX_{a}&=&\left[\omega_{a}Y_{a}-g\sqrt{2}X_{b}Y_{a}-\kappa^{\prime}_{a}X_{a}\right]dt\\
&+&dW^{h}_{x}+dW^{a}_{x},\label{eq:eqXa}\\
\nonumber dY_{a}&=&-\left[\omega_{a}X_{a}-g\sqrt{2}X_{a}X_{b}+\kappa^{\prime}_{a}Y_{a}\right]dt\\
&+&dW^{h}_{y}+dW^{a}_{y},\label{eq:eqYa}\\
dX_{b}&=&\left[\omega_{b}Y_{b}-\frac{\kappa_{b}}{2}X_{b}\right]dt
+dW^b_{x},\label{eq:eqXb}\\
\nonumber dY_{b}&=&-\left[\omega_{b}X_{b}+\frac{\kappa_{b}}{2}Y_{b}-\frac{g}{\sqrt{2}}(X_{a}^2+Y_{a}^2)\right]dt\\
&+&dW^b_y.\label{eq:eqYb}
\end{eqnarray}
Here, $dW^{i}=1/\sqrt{2}(dW_{x}^{i}+idW_{y}^{i})$ where $\xi_k^{i} dt=:dW_k^{i}$, with $k=x,y$, is the Wiener process with width 
$\sqrt{\kappa_i\bar{n}_i dt}$ and $\kappa'_{a}:=\kappa_{h}/2+\kappa_{a}/2$. For the cooling stage, $\kappa'_{a}$ is replaced by 
$\kappa_{a}/2$ and the $dW^h_x,dW^h_y$ terms vanish. 

The physical parameters used in the classical dynamical simulations are the same as those used in the quantum case. 
The equations are solved by using Mathematica 10.
\section{Results of classical simulations}\label{sec:classicalResults}
In Fig.~\ref{fig:fig11}, we present the change in the internal energy $U_a$ of the classical engine with respect to 
$\omega_{\text{eff}}$ for $\bar{n}_{h}=0.125$. The result follows and fluctuates about the semiclassical cycle in Fig.~\ref{fig:fig6}, as
expected theoretically. Similarly, the phase portrait of the classical engine in Fig.~\ref{fig:fig12} is the same with 
its quantum counterpart in Fig.~\ref{fig:fig4}. Theoretically, the quantum dynamics of $\langle \hat n_a\rangle$ and the
phase space quadratures are the same for the classical and quantum mechanical descriptions. Factorization of the two operator
expectation value $\langle \hat n_a\hat q\rangle$ to define $\omega_{\text{eff}}$ makes the semiclassical $U_a-\omega_{\text{eff}}$ cycle identical with the classical one.

Fig.~\ref{fig:fig9} shows the maximum power output of the classical engine with respect to 
$\bar{n}_{h}$. The power output of the classical  engine lies between the quantum and semiclassical models. The difference of quantum model from
the classical and the semiclassical models is more significant in the dissipated power output of the LF resonator in contrast to negligible difference
between the classical and quantum treatment of the extractable work from the
HF resonator. The differences diminish with increasing temperature of the driving noise as shown
in Fig.~\ref{fig:fig10}. In order to explain such different effects of correlations on different subsystems of the engine, we investigate the 
classical and quantum correlations 
more closely in the next section. We note that power enhancement in the coherent work extraction relative to the stochastic one has been discussed in the literature as a signature of the quantum character of certain quantum heat engines, different than
our system. Quantum coherence has been suggested as the main source of the power enhancement~\cite{uzdin_equivalence_2015}. It has been conjectured that quantum correlations (entanglement or discord) 
in multiparticle engines could play a similar role to that of coherence in single particle engines. Our two
coupled resonator set up can be envisioned as an example to verify this conjecture.
As the systems are different, we need to elaborate explicitly if and how the quantum and classical correlations emerge in our system and how they
lead to enhanced power output in our case.
\section{Quantum nature of the system}
\label{sec:classicalVSquantum}
The equations of motion for the dynamical variables of interest in our system are given in the 
Appendix~\ref{app:q-dynamics}. In this section, we will
discuss the set of equations related to the correlations contributing to the power and work output of our quantum heat engine. 

The extractable work depends on the pressure- displacement correlation
$\langle \hat n_a,\hat q\rangle$, where the correlation function for two operators 
$\hat o_1$ and $\hat o_2$
is defined as $\langle \hat o_1,\hat o_2\rangle=\langle \hat o_1\hat o_2\rangle-\langle \hat o_1\rangle\langle \hat o_2\rangle$. According to Eq.~(\ref{eq:mechPower}), the power output of the Otto cycle
of the HF resonator depends on the pressure-momentum correlation $\langle \hat n_a,\hat p\rangle$. Both the pressure-displacement and pressure-momentum correlations are
driven by the number fluctuations of the HF mode which is thermally excited by the hot reservoir. This mechanism is described by the closed
system of equations given in Appendix~\ref{app:q-dynamics}. We can find the equation of motion for $\langle \hat n_a,\hat q\rangle$ as
\begin{eqnarray}
\left(\frac{d^2}{dt^2}+2C\frac{d^2}{dt}+\omega^2\right) \langle \hat n_a,\hat q\rangle=2g\omega_b\langle(\Delta\hat n_a)^2\rangle,
\end{eqnarray}
where $C=\kappa_h+\kappa_a+\kappa_b/2$ and $\omega=\omega_b^2+C^2$.
system of equations for $\langle(\Delta\hat n_a)^2\rangle$, $\langle \hat n_a,\hat q\rangle$, and $\langle \hat n_a,\hat p\rangle$
is parallel to that of $\langle\hat n_a\rangle$, $\langle\hat q\rangle$, and $\langle\hat p\rangle$. 
 $\langle(\Delta\hat n_a)^2\rangle$ and $\langle\hat n_a\rangle$ have similar temporal profiles, with the 
 same periodicity but with different amplitudes. Accordingly, 
$\langle \hat n_a,\hat q\rangle$ exhibits a periodically-driven underdamped oscillatory motion,
which is analogous to the dynamics of $q$.
In contrast to the case of $q$, however, the weak damping condition $C<\omega$ is weakly satisfied and the correlation
dynamics are in a regime which is only slightly beyond critical damping (since $C\sim 2\kappa_a\gg\omega_b$).
\begin{figure}[!t]
	\includegraphics[width=7.5cm]{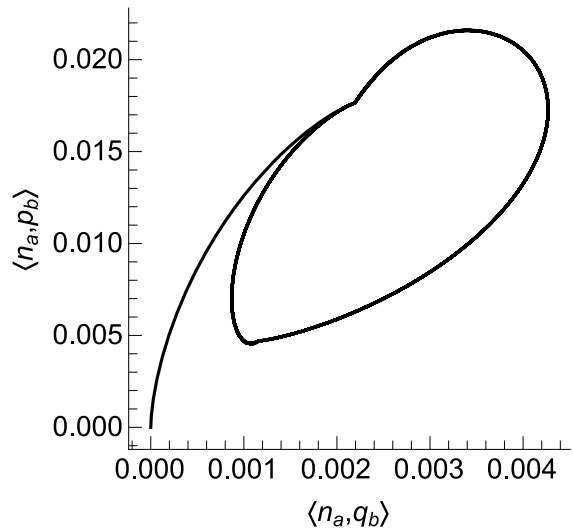}
	\caption{(Colour online) Relation between the dynamics of the
	quantum correlations $\langle \hat n_a,\hat q\rangle$ and $\langle \hat n_a,\hat q\rangle$. 
	A closed
	cycle emerges in the steady state as both correlations are slaved by the number fluctuation dynamics of HF resonator. 
	$\langle \hat n_a,\hat p\rangle$ varies over a much wider range than the $\langle \hat n_a,\hat q\rangle$.
	All the other parameters are as explained in the text.}
	\label{fig:fig13}
\end{figure}

The formal solution of the $\langle \hat n_a,\hat q\rangle$  can be written as
\begin{eqnarray}
\langle \hat n_a,\hat q\rangle =
\int_0^t dt^\prime e^{-C(t-t^\prime)}
\sin{(\omega_b(t-t^\prime))}f(t^\prime),
\end{eqnarray}
with $f(t)=2g\langle(\Delta\hat n_a)^2(t)\rangle$. 
Another difference from the $q$ dynamics, which has a resonant drive, is that the natural frequency $\omega\sim 2\kappa_a$ 
of the correlation oscillator is 
far off resonant with
the drive frequency $\omega_b$. In this case, the fast correlations adiabatically follows the slow fluctuations of the HF resonator.
We can employ integration by parts to the formal solution to show that 
\begin{eqnarray}
\langle \hat n_a,\hat q\rangle\approx \frac{g\omega_b}{2\kappa_a^2}\langle(\Delta\hat n_a)^2\rangle(1-e^{-Ct}).
\end{eqnarray}
In the long-time limit the steady state at the end of the heating and cooling cycles coincides with the exact steady state for the
continuous drive case discussed in the Appendix~\ref{app:q-steadyState}.
In terms of our parameter values, this value is too small ($\sim 10^{-3}$) to be of significance for the extractable work from the
working HF resonator. 
On the other hand, using the relation
\begin{eqnarray}
\langle \hat n_a,\hat p\rangle=\frac{1}{\omega_b}\left(\frac{d}{dt}+C\right) \langle \hat n_a,\hat q\rangle,
\end{eqnarray}
we get 
\begin{eqnarray}
\langle \hat n_a,\hat p\rangle=\frac{g}{\kappa_a}\langle(\Delta\hat n_a)^2\rangle,
\end{eqnarray}
which turns out to be an order of magnitude greater than $\langle \hat n_a,\hat q\rangle$. Their mutual dynamics are plotted
in Fig.~\ref{fig:fig13}. 
According to the Eq.~(\ref{eq:mechPower}),
instead of contributing
to the work output, $\langle \hat n_a,\hat p\rangle$ contributes to power output of the Otto cycle
of the HF mode.

\begin{figure}[!t]
	\centering
	\begin{center}
		\subfigure[]{
			\label{fig:fig14a}
			\includegraphics[width=7.5cm]{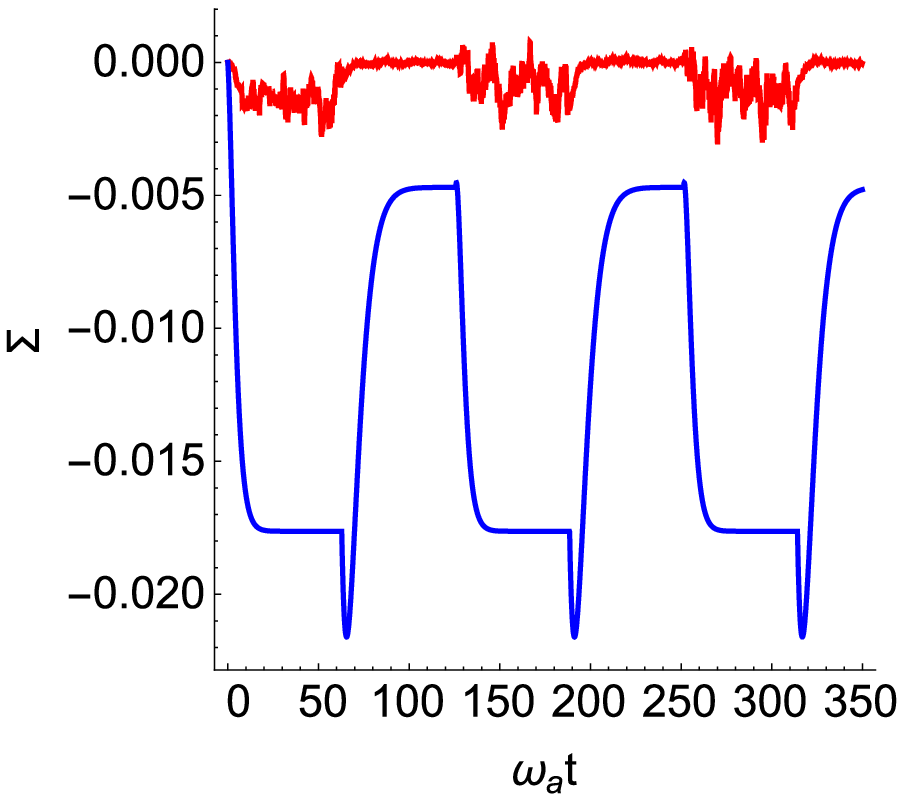}
		}
		
		\subfigure[]{
			\label{fig:fig14b}
			\includegraphics[width=7.5cm]{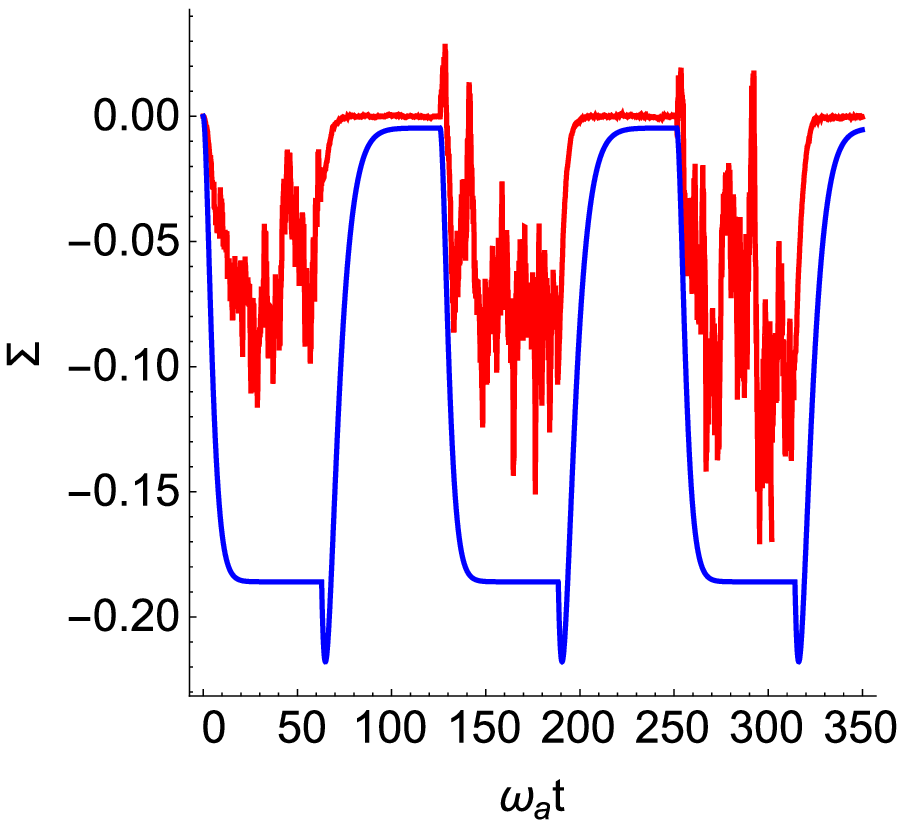}	
		}
		
	\end{center}
	\caption{(Colour online) Dynamics of the correlation function $\Sigma$ with respect to the scaled time 
	$\omega_{\text{a}}t$ for (a) $\bar{n}_{\text{h}}=0.125$ and (b) $\bar{n}_{\text{h}}=1$ calculated by 
	quantum (solid blue curves) and classical (red noisy curves) methods.
	All the other parameters are as explained in the text.}
	\label{fig:fig14}
\end{figure}

We note that a similar parameter to $\langle \hat n_a,\hat p\rangle$ (normalized by the variances of the $\hat n_a$ and $\hat p$) 
has been proposed in the context of the quality of quantum nondemolition measurements and called the signal-meter entanglement parameter~\cite{holland1990nonideal}. To make analogous  notation we use $\Sigma:=-\langle \hat n_a,\hat p\rangle$.
 It has also been used in optomechanical systems as an indicator of quantum coherence~\cite{mancini1997ponderomotive}. 
When $\Sigma<0${\color{red},} signal-meter type quantum
correlations make a positive contribution to the rate of increase of the power output from the LF resonator. The time dependence of $\Sigma$
is shown in in Fig.~\ref{fig:fig14}. It is oscillatory, adiabatically following the number fluctuations in the HF resonator, and it is always negative.
Similar oscillations in $\Sigma$ are found for a typical
optomechanical set up where the drive is coherent. It was proposed that the correlated state can be used to generate a mixture of
Cat states through conditional measurement~\cite{mancini1997ponderomotive}. In our case, coherence and correlations are driven by an
incoherent drive. 

We have also calculated the classical correlation $\Sigma_{\text{class}}=-\langle \hat n_a,\hat p\rangle_{\text{class}}$.
Here $\langle .\rangle_{\text{class}}$ indicates averaging over classical trajectories.
The results are plotted in Fig.~\ref{fig:fig14}. This shows that there is a difference between the amplitudes of the quantum and classical correlations. The quantum model
predicts a faster growth rate and higher maximum value for the power output from the LF resonator relative to the classical model. Both models
predict more powerful and faster engines relative to the semiclassical model.


\section{Conclusions}\label{sec:conc}
In summary, we  proposed and investigated a pair of superconducting transmission line resonators  
with an optomechanical-like
interaction as a quantum
heat engine. The high frequency (HF) resonator is periodically coupled to an effective hot bath while both resonators are 
coupled to continuous effective cold baths.
We found the emergence of a limit cycle in the open system dynamics and identified that
the HF resonator mode undergoes an effective Otto cycle. The electromagnetic field mode of the HF resonator acts as the working 
substance while the mode of the LF one
acts as the piston that can be used for coherent power extraction. The superconducting resonator engine serves as an electrical
analog of a mechanical piston engine. We found that a piston-like motion can be identified in the phase space 
by using the phase portraits and Wigner function plots.
We numerically verified, by calculating the second order coherence 
function, that the working resonator remains thermal while the ``piston'' resonator is ``charged"
to a thermal coherent state, which is a thermodynamically non-passive state that can be used to harvest
useful work~\cite{gelbwaser-klimovsky_work_2015}.

Extractable work and efficiency has been calculated from the T-S diagram and from the internal energy diagram.
We found that pressure-displacement correlations could in principle contribute to the enhancement of extractable
work. However, the effect is negligible. A practical figure of merit for the piston engine performance is the dissipative
power from the piston subsystem. When we evaluate it, we found that pressure-momentum correlations (also known as signal-meter
type correlations) contribute
significantly. These correlations are driven by the 
variance of the working resonator, which is driven by the thermal fluctuations of the hot reservoir. 
By this way, in addition to the mean number of excitations in the hot bath, its fluctuations can
be harvested through the dissipative power of the piston resonator.
This is not a universal conclusion. In other optomechanical quantum piston engines quantum
correlations could lead to a power increase only in limited parameter regimes~\cite{mari_quantum_2015} or could even be harmful to
the output power~\cite{roulet_autonomous_2017}.
We compared the classical, semiclassical, and quantum engine descriptions and concluded that our engine
is inevitably a genuine quantum heat engine with enhanced power output due to 
quantum correlations so that it could be used as a test bed to explore quantum effects in
quantum engine cycles.

Here we have only considered classical noise drives, however our framework is applicable to the case of quantum noise drives as well. According to our analytical results, 
we conclude that 
the quantum enhancement in the power output can be further
increased by using quantum drives instead of classical drives. For example, using a squeezed thermal
noise, number fluctuations could be increased to enlarge the pressure-displacement or pressure-momentum correlations to
make the quantum enhancement in the work or power output more significant relative to classical engines~\cite{rosnagel_nanoscale_2014,klaers_squeezed_2017}.

Our results could be practically significant for realization of 
compact, on-chip, scalable, electronic realizations of genuine 
quantum heat engines with quantum coherence and correlations advantages relative to their classical counterparts. It can also fundamentally
be used as a testbed to explore the quantum-to-classical transition of heat engines and to illuminate the role of quantum correlations in
the operation of quantum heat engines.
\begin{acknowledgements}
We acknowledge Pol Forn-Diaz, Kay Brandner, and Mika Sillanp{\"a}{\"a}  for useful discussions. 
A.~\"U.~C.~H.~acknowledges support from the Villum Foundation through a postdoctoral block stipend. 
\"O.~E.~M.~and A.~\"U.~C.~H.~acknowledge 
the support by Lockheed Martin Corporation and Ko\c{c} University Research Agreement.
C.~M.~W.~acknowledges the support by Lockheed Martin Corporation and University of Waterloo 
Research Agreement.
\end{acknowledgements}
\appendix
\section{Quantum dynamics of the model system}
\label{app:q-dynamics}
A ``gobal" master equation, also called the ``dressed state master equation (DSME)", which is applicable to a system with arbitrarily strong optomechanical
coupling has been derived in Ref.~\cite{hu_quantum_2015} and expressed in the Schro\"{o}dinger
picture as $(\hbar=1,k_B=1)$
\begin{eqnarray}\label{eq:globalmaster}
\dot{\hat{\rho}}&=&-i[\hat{H}_{\text{sys}},\hat{\rho}] \\ \nonumber
&+&\kappa_{a}(\bar{n}_{a}+1)D[\hat{a}]+\kappa_{a}\bar{n}_{a}D[\hat{a}^{\dagger}]\\ \nonumber 
&+&\kappa_{\text{h}}(t)(\bar{n}_{\text{h}}+1)D[\hat{a}]+\kappa_{h}(t)\bar{n}_{\text{h}}D[\hat{a}^{\dagger}]
\\ \nonumber 
&+&\alpha^2\kappa_{\text{d}}D[\hat{a}^{\dagger}\hat{a}]
\\ \nonumber 
&+&\kappa_{b}(\bar{n}_{b}+1)D[\hat{b}-\alpha\hat{a}^{\dagger}\hat{a}]
+\kappa_{b}\bar{n}_{b}D[\hat{b}^{\dagger}-\alpha\hat{a}^{\dagger}\hat{a}],
\end{eqnarray}
where $\alpha=g/\omega_b$ characterizes the ``non-local' effects of the reservoirs 
and $\kappa_{\text{d}}=4\kappa_bT_b/\omega_b$ is the dephasing rate of the HF resonator
mode. This expression is obtained under the assumption that the bath of the LF resonator has an
Ohmic spectral density. The interaction of the HF resonator mode with the baths attached to it remain
approximately local as $\omega_a\gg\omega_b$ so that ``phonon" side modes do not change
the spectral densities of the baths of the HF resonator. This requires the spectral densities to 
be slowly varying near the resonance frequency $\omega_a$. These are consistent with the conditions we employ on the noise drives applied to the resonators.

Equations of motions for the relevant dynamical observables of our system are 
determined from the master equation and given by
\begin{eqnarray}
&\frac{d}{dt}&\langle\hat n_a\rangle=A-B\langle\hat n_a\rangle, \\
&\frac{d}{dt}&\langle\hat p\rangle=-\omega_b\langle\hat q\rangle-\frac{\kappa_b}{2}\langle\hat p\rangle+2g\langle\hat n_a\rangle,\\
&\frac{d}{dt}&\langle\hat q\rangle=\omega_b\langle\hat p\rangle-\frac{\kappa_b}{2}\langle\hat q\rangle+\alpha\kappa_b\langle\hat n_a\rangle,\\
&\frac{d}{dt}&\langle(\Delta\hat n_a)^2\rangle=A+(2A+B)\langle\hat n_a\rangle-2B\langle(\Delta\hat n_a)^2\rangle, \\
&\frac{d}{dt}&\langle \hat n_a,\hat p\rangle=-C\langle \hat n_a,\hat p\rangle-\omega_b\langle 
\hat n_a,\hat q\rangle+2g\langle(\Delta\hat n_a)^2\rangle,\\
&\frac{d}{dt}&\langle \hat n_a,\hat q\rangle=-C\langle \hat n_a,\hat q\rangle+\omega_b\langle 
\hat n_a,\hat p\rangle+\alpha\kappa_b\langle(\Delta\hat n_a)^2\rangle,\\
&\frac{d}{dt}&\langle \hat n_b\rangle=-\kappa_b(\langle \hat n_b\rangle-\bar n_b)+
g(\langle \hat n_a,\hat p\rangle+\langle \hat n_a\rangle\langle\hat p\rangle)\nonumber\\
&&+\alpha\frac{\kappa_b}{2}(\langle \hat n_a,\hat q\rangle+\langle\hat n_a\rangle\langle\hat q\rangle),
\end{eqnarray}
where $\hat q=\hat b+\hat b^\dag$, $\hat p=i(\hat b-\hat b^\dag)$,
$A=\kappa_a\bar n_a+\kappa_h\bar n_h$, $B=\kappa_a+\kappa_h$, and $C=B+\kappa_b/2$. Correlation function between the two
dynamical observables with the operators $\hat o_1$ and $ \hat o_2$ are denoted by
$\langle \hat o_1,\hat o_2\rangle =\langle \hat o_1\hat o_2\rangle-\langle \hat o_1\rangle\langle \hat o_2\rangle$. The terms with parameter $\alpha$ are due to the non-local effects of the
reservoir-system interactions. In the range of our interest, $g\le\omega_b$ we did not find any significant effects due to them, except a negligible increase in $\langle \hat n_a,\hat q\rangle$. We use the local
master equation and the corresponding equations of motion by taking $\alpha=0$ unless otherwise
noted in the manuscript.

We note that the closed dynamics that we have found 
for the set of thermodynamically relevant observables cannot be found for all dynamical variables 
of the system. The optomechanical Hamiltonian has a special symmetry in which the number operator
of the HF resonator mode is a constant of motion. The steady state of the HF resonator and the associated
local observables for the HF mode are independent of the optomechanical coupling coefficient $g$~\cite{bernad_partly_2015}.
The first three equations of motion form a closed set of dynamics for the $q-p$ phase space trajectories,
driven by $\langle\hat n_a\rangle$. The second set of three equations form another, inpendent, 
closed set of dynamics
for the ``pressure-displacement", $\langle \hat n_a,\hat q\rangle$, and ``pressure-momentum", $\langle \hat n_a,\hat p\rangle$ correlations, 
driven by the quantum fluctuations
$\langle(\Delta\hat n_a)^2\rangle$. The baths of the HF resonator mode prepare the pressure 
($\sim\langle \hat n_a\rangle$ and quantum noise drives for these independent dynamics.
Both dynamics contribute to the evolution of the excitations in the LF mode given in the last equation.
\section{Steady state of the quantum dynamics under continuous heating}
\label{app:q-steadyState}
While the dynamics of the LF mode is different under pulsed thermal drive, the HF resonator reaches approximately its steady state 
identical with a continuous drive case, for the pulse duration is longer than its thermalization transition time. In order to
highlight the effect of pulsed drive relative to continuous drive on the LF mode dynamics, and to give steady state
values of the HF dynamics as well, we provide the solutions of the quantum dynamical equations in the 
Appendix~\ref{app:q-dynamics} in the long time limit below and note some common characteristic behaviors.
\begin{eqnarray}
\langle\hat n_a\rangle_{\text{SS}}&=&\frac{A}{B}=\frac{\kappa_a\bar n_a+\kappa_h\bar n_h}{\kappa_a+\kappa_h},\\
\langle\hat q\rangle_{\text{SS}} &=& \frac{8g\omega_b+4\alpha\kappa_b^2}{4\omega_b^2+\kappa_b^2}\langle\hat n_a\rangle_{\text{SS}},\\
\langle\hat p\rangle_{\text{SS}} &=&\frac{4g\kappa_b-4\alpha\kappa_b\omega_b}{4\omega_b^2+\kappa_b^2}\langle\hat n_a\rangle_{\text{SS}}=0,\\
\langle(\Delta\hat n_a)^2\rangle_{\text{SS}} &=&\frac{A(A+B)}{B^2},\\
\langle \hat n_a,\hat p\rangle_{\text{SS}}&=&\frac{2gC-\alpha\kappa_b\omega_b}{\omega_b^2+C^2}\langle(\Delta\hat n_a)^2\rangle_{\text{SS}},\\
&=&\frac{2gB}{\omega_b^2+C^2}\langle(\Delta\hat n_a)^2\rangle_{\text{SS}},\\
\langle \hat n_a,\hat q\rangle_{\text{SS}}&=&\frac{2g\omega_b+\alpha\kappa_bC}{\omega_b^2+C^2}\langle(\Delta\hat n_a)^2\rangle_{\text{SS}},\\
\langle \hat n_b\rangle_{\text{SS}}&=&\bar n_b
+\frac{g}{\kappa_b}(\langle \hat n_a,\hat p\rangle_{\text{SS}}
+\langle\hat n_a\rangle_{\text{SS}}\langle\hat p\rangle_{\text{SS}})\nonumber\\
&+&\frac{\alpha}{2}(\langle \hat n_a,\hat q\rangle+\langle\hat n_a\rangle\langle\hat q\rangle)
\end{eqnarray}
According to our parameters we have $\kappa_a=\kappa_h$ so that $\langle \hat n_a\rangle_{\text{SS}} = (\bar n_c+\bar n_h)/2$. In
addition, the variance becomes $\langle(\Delta\hat n_a)^2\rangle_{\text{SS}}\sim (\bar n_a+\bar n_h)(\bar n_a+\bar n_h+2)/4$.
Exact solution
of the pulsed drive case yields these value to a very good approximation, conforming that the pulse duration is sufficiently
long to thermalize the HF resonator. The steady state values of the LF resonator gives smaller $\langle \hat n_b\rangle_{\text{SS}}$
in the case of continuous drive. A common behavior under both the continuous or pulsed drive is 
the quadratic increase of $\langle \hat n_b\rangle_{\text{SS}}$ with $\bar n_h$. Both the correlation term $\langle \hat n_a,\hat p\rangle_{\text{SS}}$
and the factorized semi-classical term  $\langle\hat n_a\rangle_{\text{SS}}\langle\hat p\rangle_{\text{SS}}$  increase quadratically with  $\bar n_h$.
This behavior is translated to the power output of the engine, which is proportional to $P\propto \langle \hat n_b\rangle_{\text{SS}}-\bar n_b$.\\
\section{Steady state of the electric displacement under periodic heating}
\label{app:q-GenSteadyState}
The equation motion of $q$
\begin{eqnarray}
\ddot{q}+\kappa_b\dot{q}+\omega^2q=2\omega_bg\langle n_a(t)\rangle,
\end{eqnarray}
with $\omega^2=\omega_b^2+\kappa_b^2/4$,
can be solved in steady state by approximating the $\langle\hat n_a\rangle$ by a square wave with period $T=2\pi/\omega_b$. Transforming 
$q\rightarrow q=Q+q_0$ we
write
\begin{eqnarray}
\ddot{Q}+\kappa_b\dot{Q}+\omega^2Q=2\omega_bg\langle F_a(t)\rangle,
\end{eqnarray}
where $F(t)$ is made to be periodically alternating between $\pm (\bar n_h-\bar n_c)/4$ by taking
\begin{eqnarray}
q_0=\frac{2\omega_bg}{\omega^2}\left(\frac{\bar n_h+3\bar n_c}{4}\right).
\end{eqnarray}
First harmonic at frequency $\omega_b$ is then given by
\begin{eqnarray}
Q=\frac{4}{\pi}\frac{2\omega_bg}{\sqrt{\kappa_b^2\left(\omega_b^2+\frac{\kappa_b^2}{16}\right)}}
\left(\frac{\bar n_h-\bar n_c}{4}\right)\sin{(\omega_bt+\alpha)},
\end{eqnarray}
where 
\begin{eqnarray}
\alpha=\arctan{\left(\frac{\kappa_b\omega_b}{\omega_b^2-\omega^2}\right)}.
\end{eqnarray}
\bibliographystyle{apsrev4-1}
\bibliography{cQEDeng2017}

\begin{thebibliography}{58}%
\makeatletter
\providecommand \@ifxundefined [1]{%
 \@ifx{#1\undefined}
}%
\providecommand \@ifnum [1]{%
 \ifnum #1\expandafter \@firstoftwo
 \else \expandafter \@secondoftwo
 \fi
}%
\providecommand \@ifx [1]{%
 \ifx #1\expandafter \@firstoftwo
 \else \expandafter \@secondoftwo
 \fi
}%
\providecommand \natexlab [1]{#1}%
\providecommand \enquote  [1]{``#1''}%
\providecommand \bibnamefont  [1]{#1}%
\providecommand \bibfnamefont [1]{#1}%
\providecommand \citenamefont [1]{#1}%
\providecommand \href@noop [0]{\@secondoftwo}%
\providecommand \href [0]{\begingroup \@sanitize@url \@href}%
\providecommand \@href[1]{\@@startlink{#1}\@@href}%
\providecommand \@@href[1]{\endgroup#1\@@endlink}%
\providecommand \@sanitize@url [0]{\catcode `\\12\catcode `\$12\catcode
  `\&12\catcode `\#12\catcode `\^12\catcode `\_12\catcode `\%12\relax}%
\providecommand \@@startlink[1]{}%
\providecommand \@@endlink[0]{}%
\providecommand \url  [0]{\begingroup\@sanitize@url \@url }%
\providecommand \@url [1]{\endgroup\@href {#1}{\urlprefix }}%
\providecommand \urlprefix  [0]{URL }%
\providecommand \Eprint [0]{\href }%
\providecommand \doibase [0]{http://dx.doi.org/}%
\providecommand \selectlanguage [0]{\@gobble}%
\providecommand \bibinfo  [0]{\@secondoftwo}%
\providecommand \bibfield  [0]{\@secondoftwo}%
\providecommand \translation [1]{[#1]}%
\providecommand \BibitemOpen [0]{}%
\providecommand \bibitemStop [0]{}%
\providecommand \bibitemNoStop [0]{.\EOS\space}%
\providecommand \EOS [0]{\spacefactor3000\relax}%
\providecommand \BibitemShut  [1]{\csname bibitem#1\endcsname}%
\let\auto@bib@innerbib\@empty
\bibitem [{\citenamefont {Kieu}()}]{kieu2004second}%
  \BibitemOpen
  \bibfield  {author} {\bibinfo {author} {\bibfnamefont {T.~D.}\ \bibnamefont
  {Kieu}},\ }\bibfield  {title} {\emph {\bibinfo {title} {The second law,
  maxwell's demon, and work derivable from quantum heat engines},\ }}\href
  {https://doi.org/10.1103/PhysRevLett.93.140403} {\bibfield  {journal}
  {\bibinfo  {journal} {Phys. Rev. Lett.}\ }\textbf {\bibinfo {volume} {93}},\
  \bibinfo {pages} {140403}}\BibitemShut {NoStop}%
\bibitem [{\citenamefont {Quan}\ \emph {et~al.}(2007)\citenamefont {Quan},
  \citenamefont {Liu}, \citenamefont {Sun},\ and\ \citenamefont
  {Nori}}]{quan_quantum_2007}%
  \BibitemOpen
  \bibfield  {author} {\bibinfo {author} {\bibfnamefont {H.~T.}\ \bibnamefont
  {Quan}}, \bibinfo {author} {\bibfnamefont {Y.-x.}\ \bibnamefont {Liu}},
  \bibinfo {author} {\bibfnamefont {C.~P.}\ \bibnamefont {Sun}}, \ and\
  \bibinfo {author} {\bibfnamefont {F.}~\bibnamefont {Nori}},\ }\bibfield
  {title} {\emph {\bibinfo {title} {Quantum thermodynamic cycles and quantum
  heat engines},\ }}\href {\doibase 10.1103/PhysRevE.76.031105} {\bibfield
  {journal} {\bibinfo  {journal} {Phys. Rev. E}\ }\textbf {\bibinfo {volume}
  {76}},\ \bibinfo {pages} {031105} (\bibinfo {year} {2007})}\BibitemShut
  {NoStop}%
\bibitem [{\citenamefont {Quan}\ \emph {et~al.}(2005)\citenamefont {Quan},
  \citenamefont {Zhang},\ and\ \citenamefont {Sun}}]{quan_quantum_2005}%
  \BibitemOpen
  \bibfield  {author} {\bibinfo {author} {\bibfnamefont {H.~T.}\ \bibnamefont
  {Quan}}, \bibinfo {author} {\bibfnamefont {P.}~\bibnamefont {Zhang}}, \ and\
  \bibinfo {author} {\bibfnamefont {C.~P.}\ \bibnamefont {Sun}},\ }\bibfield
  {title} {\emph {\bibinfo {title} {Quantum heat engine with multilevel quantum
  systems},\ }}\href {\doibase 10.1103/PhysRevE.72.056110} {\bibfield
  {journal} {\bibinfo  {journal} {Phys. Rev. E}\ }\textbf {\bibinfo {volume}
  {72}},\ \bibinfo {pages} {056110} (\bibinfo {year} {2005})}\BibitemShut
  {NoStop}%
\bibitem [{\citenamefont {Abah}\ \emph {et~al.}(2012)\citenamefont {Abah},
  \citenamefont {Ro{\ss}nagel}, \citenamefont {Jacob}, \citenamefont {Deffner},
  \citenamefont {Schmidt-Kaler}, \citenamefont {Singer},\ and\ \citenamefont
  {Lutz}}]{abah_single-ion_2012}%
  \BibitemOpen
  \bibfield  {author} {\bibinfo {author} {\bibfnamefont {O.}~\bibnamefont
  {Abah}}, \bibinfo {author} {\bibfnamefont {J.}~\bibnamefont {Ro{\ss}nagel}},
  \bibinfo {author} {\bibfnamefont {G.}~\bibnamefont {Jacob}}, \bibinfo
  {author} {\bibfnamefont {S.}~\bibnamefont {Deffner}}, \bibinfo {author}
  {\bibfnamefont {F.}~\bibnamefont {Schmidt-Kaler}}, \bibinfo {author}
  {\bibfnamefont {K.}~\bibnamefont {Singer}}, \ and\ \bibinfo {author}
  {\bibfnamefont {E.}~\bibnamefont {Lutz}},\ }\bibfield  {title} {\emph
  {\bibinfo {title} {Single-{Ion} {Heat} {Engine} at {Maximum} {Power}},\
  }}\href {\doibase 10.1103/PhysRevLett.109.203006} {\bibfield  {journal}
  {\bibinfo  {journal} {Phys. Rev. Lett.}\ }\textbf {\bibinfo {volume} {109}},\
  \bibinfo {pages} {203006} (\bibinfo {year} {2012})}\BibitemShut {NoStop}%
\bibitem [{\citenamefont {Bergenfeldt}\ \emph {et~al.}(2014)\citenamefont
  {Bergenfeldt}, \citenamefont {Samuelsson}, \citenamefont {Sothmann},
  \citenamefont {Flindt},\ and\ \citenamefont
  {B{\"u}ttiker}}]{bergenfeldt2014hybrid}%
  \BibitemOpen
  \bibfield  {author} {\bibinfo {author} {\bibfnamefont {C.}~\bibnamefont
  {Bergenfeldt}}, \bibinfo {author} {\bibfnamefont {P.}~\bibnamefont
  {Samuelsson}}, \bibinfo {author} {\bibfnamefont {B.}~\bibnamefont
  {Sothmann}}, \bibinfo {author} {\bibfnamefont {C.}~\bibnamefont {Flindt}}, \
  and\ \bibinfo {author} {\bibfnamefont {M.}~\bibnamefont {B{\"u}ttiker}},\
  }\bibfield  {title} {\emph {\bibinfo {title} {Hybrid microwave-cavity heat
  engine},\ }}\href {https://doi.org/10.1103/PhysRevLett.112.076803} {\bibfield
   {journal} {\bibinfo  {journal} {Phys. Rev. Lett.}\ }\textbf {\bibinfo
  {volume} {112}},\ \bibinfo {pages} {076803} (\bibinfo {year}
  {2014})}\BibitemShut {NoStop}%
\bibitem [{\citenamefont {Henrich}\ \emph {et~al.}(2007)\citenamefont
  {Henrich}, \citenamefont {Mahler},\ and\ \citenamefont
  {Michel}}]{henrich_driven_2007}%
  \BibitemOpen
  \bibfield  {author} {\bibinfo {author} {\bibfnamefont {M.~J.}\ \bibnamefont
  {Henrich}}, \bibinfo {author} {\bibfnamefont {G.}~\bibnamefont {Mahler}}, \
  and\ \bibinfo {author} {\bibfnamefont {M.}~\bibnamefont {Michel}},\
  }\bibfield  {title} {\emph {\bibinfo {title} {Driven spin systems as quantum
  thermodynamic machines: {Fundamental} limits},\ }}\href {\doibase
  10.1103/PhysRevE.75.051118} {\bibfield  {journal} {\bibinfo  {journal} {Phys.
  Rev. E}\ }\textbf {\bibinfo {volume} {75}},\ \bibinfo {pages} {051118}
  (\bibinfo {year} {2007})}\BibitemShut {NoStop}%
\bibitem [{\citenamefont {Uzdin}\ and\ \citenamefont
  {Kosloff}(2014)}]{uzdin_multilevel_2014}%
  \BibitemOpen
  \bibfield  {author} {\bibinfo {author} {\bibfnamefont {R.}~\bibnamefont
  {Uzdin}}\ and\ \bibinfo {author} {\bibfnamefont {R.}~\bibnamefont
  {Kosloff}},\ }\bibfield  {title} {\emph {\bibinfo {title} {The multilevel
  four-stroke swap engine and its environment},\ }}\href {\doibase
  10.1088/1367-2630/16/9/095003} {\bibfield  {journal} {\bibinfo  {journal}
  {New J. Phys.}\ }\textbf {\bibinfo {volume} {16}},\ \bibinfo {pages} {095003}
  (\bibinfo {year} {2014})}\BibitemShut {NoStop}%
\bibitem [{\citenamefont {Scully}\ \emph {et~al.}(2003)\citenamefont {Scully},
  \citenamefont {Zubairy}, \citenamefont {Agarwal},\ and\ \citenamefont
  {Walther}}]{scully2003extracting}%
  \BibitemOpen
  \bibfield  {author} {\bibinfo {author} {\bibfnamefont {M.~O.}\ \bibnamefont
  {Scully}}, \bibinfo {author} {\bibfnamefont {M.~S.}\ \bibnamefont {Zubairy}},
  \bibinfo {author} {\bibfnamefont {G.~S.}\ \bibnamefont {Agarwal}}, \ and\
  \bibinfo {author} {\bibfnamefont {H.}~\bibnamefont {Walther}},\ }\bibfield
  {title} {\emph {\bibinfo {title} {Extracting work from a single heat bath via
  vanishing quantum coherence},\ }}\href
  {https://doi.org/10.1126/science.1078955} {\bibfield  {journal} {\bibinfo
  {journal} {Science}\ }\textbf {\bibinfo {volume} {299}},\ \bibinfo {pages}
  {862} (\bibinfo {year} {2003})}\BibitemShut {NoStop}%
\bibitem [{\citenamefont {Scully}\ \emph {et~al.}(2011)\citenamefont {Scully},
  \citenamefont {Chapin}, \citenamefont {Dorfman}, \citenamefont {Kim},\ and\
  \citenamefont {Svidzinsky}}]{scully2011quantum}%
  \BibitemOpen
  \bibfield  {author} {\bibinfo {author} {\bibfnamefont {M.~O.}\ \bibnamefont
  {Scully}}, \bibinfo {author} {\bibfnamefont {K.~R.}\ \bibnamefont {Chapin}},
  \bibinfo {author} {\bibfnamefont {K.~E.}\ \bibnamefont {Dorfman}}, \bibinfo
  {author} {\bibfnamefont {M.~B.}\ \bibnamefont {Kim}}, \ and\ \bibinfo
  {author} {\bibfnamefont {A.}~\bibnamefont {Svidzinsky}},\ }\bibfield  {title}
  {\emph {\bibinfo {title} {Quantum heat engine power can be increased by
  noise-induced coherence},\ }}\href {https://doi.org/10.1073/pnas.1110234108}
  {\bibfield  {journal} {\bibinfo  {journal} {Proc. Natl. Acad. Sci.}\ }\textbf
  {\bibinfo {volume} {108}},\ \bibinfo {pages} {15097} (\bibinfo {year}
  {2011})}\BibitemShut {NoStop}%
\bibitem [{\citenamefont {Scully}(2010)}]{scully2010quantum}%
  \BibitemOpen
  \bibfield  {author} {\bibinfo {author} {\bibfnamefont {M.~O.}\ \bibnamefont
  {Scully}},\ }\bibfield  {title} {\emph {\bibinfo {title} {Quantum photocell:
  Using quantum coherence to reduce radiative recombination and increase
  efficiency},\ }}\href {https://doi.org/10.1103/PhysRevLett.104.207701}
  {\bibfield  {journal} {\bibinfo  {journal} {Phys. Rev. Lett.}\ }\textbf
  {\bibinfo {volume} {104}},\ \bibinfo {pages} {207701} (\bibinfo {year}
  {2010})}\BibitemShut {NoStop}%
\bibitem [{\citenamefont {Brunner}\ \emph {et~al.}(2012)\citenamefont
  {Brunner}, \citenamefont {Linden}, \citenamefont {Popescu},\ and\
  \citenamefont {Skrzypczyk}}]{brunner_virtual_2012}%
  \BibitemOpen
  \bibfield  {author} {\bibinfo {author} {\bibfnamefont {N.}~\bibnamefont
  {Brunner}}, \bibinfo {author} {\bibfnamefont {N.}~\bibnamefont {Linden}},
  \bibinfo {author} {\bibfnamefont {S.}~\bibnamefont {Popescu}}, \ and\
  \bibinfo {author} {\bibfnamefont {P.}~\bibnamefont {Skrzypczyk}},\ }\bibfield
   {title} {\emph {\bibinfo {title} {Virtual qubits, virtual temperatures, and
  the foundations of thermodynamics},\ }}\href {\doibase
  10.1103/PhysRevE.85.051117} {\bibfield  {journal} {\bibinfo  {journal} {Phys.
  Rev. E}\ }\textbf {\bibinfo {volume} {85}},\ \bibinfo {pages} {051117}
  (\bibinfo {year} {2012})}\BibitemShut {NoStop}%
\bibitem [{\citenamefont {Manzano}\ \emph {et~al.}(2016)\citenamefont
  {Manzano}, \citenamefont {Galve}, \citenamefont {Zambrini},\ and\
  \citenamefont {Parrondo}}]{manzano_entropy_2016}%
  \BibitemOpen
  \bibfield  {author} {\bibinfo {author} {\bibfnamefont {G.}~\bibnamefont
  {Manzano}}, \bibinfo {author} {\bibfnamefont {F.}~\bibnamefont {Galve}},
  \bibinfo {author} {\bibfnamefont {R.}~\bibnamefont {Zambrini}}, \ and\
  \bibinfo {author} {\bibfnamefont {J.~M.~R.}\ \bibnamefont {Parrondo}},\
  }\bibfield  {title} {\emph {\bibinfo {title} {Entropy production and
  thermodynamic power of the squeezed thermal reservoir},\ }}\href {\doibase
  10.1103/PhysRevE.93.052120} {\bibfield  {journal} {\bibinfo  {journal} {Phys.
  Rev. E}\ }\textbf {\bibinfo {volume} {93}},\ \bibinfo {pages} {052120}
  (\bibinfo {year} {2016})}\BibitemShut {NoStop}%
\bibitem [{\citenamefont {Song}\ \emph {et~al.}(2016)\citenamefont {Song},
  \citenamefont {Singh}, \citenamefont {Zhang}, \citenamefont {Zhang},\ and\
  \citenamefont {Meystre}}]{song_one_2016}%
  \BibitemOpen
  \bibfield  {author} {\bibinfo {author} {\bibfnamefont {Q.}~\bibnamefont
  {Song}}, \bibinfo {author} {\bibfnamefont {S.}~\bibnamefont {Singh}},
  \bibinfo {author} {\bibfnamefont {K.}~\bibnamefont {Zhang}}, \bibinfo
  {author} {\bibfnamefont {W.}~\bibnamefont {Zhang}}, \ and\ \bibinfo {author}
  {\bibfnamefont {P.}~\bibnamefont {Meystre}},\ }\bibfield  {title} {\emph
  {\bibinfo {title} {One atom and one photon - the simplest polaritonic heat
  engine},\ }}\href {http://arxiv.org/abs/1607.00119} {\bibfield  {journal}
  {\bibinfo  {journal} {arXiv:1607.00119 [quant-ph]}\ } (\bibinfo {year}
  {2016})},\ \bibinfo {note} {arXiv: 1607.00119}\BibitemShut {NoStop}%
\bibitem [{\citenamefont {Altintas}\ \emph {et~al.}(2015)\citenamefont
  {Altintas}, \citenamefont {Hardal},\ and\ \citenamefont
  {M{\"u}stecapl{\i}o{\u{g}}lu}}]{altintas2015rabi}%
  \BibitemOpen
  \bibfield  {author} {\bibinfo {author} {\bibfnamefont {F.}~\bibnamefont
  {Altintas}}, \bibinfo {author} {\bibfnamefont {A.~{\"U}.~C.}\ \bibnamefont
  {Hardal}}, \ and\ \bibinfo {author} {\bibfnamefont {{\"O}.~E.}\ \bibnamefont
  {M{\"u}stecapl{\i}o{\u{g}}lu}},\ }\bibfield  {title} {\emph {\bibinfo {title}
  {Rabi model as a quantum coherent heat engine: From quantum biology to
  superconducting circuits},\ }}\href
  {https://doi.org/10.1103/PhysRevA.91.023816} {\bibfield  {journal} {\bibinfo
  {journal} {Phys. Rev. A}\ }\textbf {\bibinfo {volume} {91}},\ \bibinfo
  {pages} {023816} (\bibinfo {year} {2015})}\BibitemShut {NoStop}%
\bibitem [{\citenamefont {Ivanchenko}(2015)}]{ivanchenko_quantum_2015}%
  \BibitemOpen
  \bibfield  {author} {\bibinfo {author} {\bibfnamefont {E.~A.}\ \bibnamefont
  {Ivanchenko}},\ }\bibfield  {title} {\emph {\bibinfo {title} {Quantum {Otto}
  cycle efficiency on coupled qudits},\ }}\href {\doibase
  10.1103/PhysRevE.92.032124} {\bibfield  {journal} {\bibinfo  {journal} {Phys.
  Rev. E}\ }\textbf {\bibinfo {volume} {92}},\ \bibinfo {pages} {032124}
  (\bibinfo {year} {2015})}\BibitemShut {NoStop}%
\bibitem [{\citenamefont {Altintas}\ \emph {et~al.}(2014)\citenamefont
  {Altintas}, \citenamefont {Hardal},\ and\ \citenamefont
  {M{\"u}stecapl{\i}og̃lu}}]{altintas2014quantum}%
  \BibitemOpen
  \bibfield  {author} {\bibinfo {author} {\bibfnamefont {F.}~\bibnamefont
  {Altintas}}, \bibinfo {author} {\bibfnamefont {A.~{\"U}.~C.}\ \bibnamefont
  {Hardal}}, \ and\ \bibinfo {author} {\bibfnamefont {{\"O}.~E.}\ \bibnamefont
  {M{\"u}stecapl{\i}og̃lu}},\ }\bibfield  {title} {\emph {\bibinfo {title}
  {Quantum correlated heat engine with spin squeezing},\ }}\href
  {https://doi.org/10.1103/PhysRevE.90.032102} {\bibfield  {journal} {\bibinfo
  {journal} {Phys. Rev. E}\ }\textbf {\bibinfo {volume} {90}},\ \bibinfo
  {pages} {032102} (\bibinfo {year} {2014})}\BibitemShut {NoStop}%
\bibitem [{\citenamefont {Da{\u g}}\ \emph {et~al.}(2016)\citenamefont {Da{\u
  g}}, \citenamefont {Niedenzu}, \citenamefont {M{\"u}stecaplıo{\u g}lu},\
  and\ \citenamefont {Kurizki}}]{dag_multiatom_2016}%
  \BibitemOpen
  \bibfield  {author} {\bibinfo {author} {\bibfnamefont {C.~B.}\ \bibnamefont
  {Da{\u g}}}, \bibinfo {author} {\bibfnamefont {W.}~\bibnamefont {Niedenzu}},
  \bibinfo {author} {\bibfnamefont {{\"O}.~E.}\ \bibnamefont
  {M{\"u}stecaplıo{\u g}lu}}, \ and\ \bibinfo {author} {\bibfnamefont
  {G.}~\bibnamefont {Kurizki}},\ }\bibfield  {title} {\emph {\bibinfo {title}
  {Multiatom {Quantum} {Coherences} in {Micromasers} as {Fuel} for {Thermal}
  and {Nonthermal} {Machines}},\ }}\href {\doibase 10.3390/e18070244}
  {\bibfield  {journal} {\bibinfo  {journal} {Entropy}\ }\textbf {\bibinfo
  {volume} {18}},\ \bibinfo {pages} {244} (\bibinfo {year} {2016})}\BibitemShut
  {NoStop}%
\bibitem [{\citenamefont {T{\"u}rkpen{\c{c}}e}\ and\ \citenamefont
  {M{\"u}stecapl{\i}o{\u{g}}lu}(2016)}]{turkpencce2016quantum}%
  \BibitemOpen
  \bibfield  {author} {\bibinfo {author} {\bibfnamefont {D.}~\bibnamefont
  {T{\"u}rkpen{\c{c}}e}}\ and\ \bibinfo {author} {\bibfnamefont {{\"O}.~E.}\
  \bibnamefont {M{\"u}stecapl{\i}o{\u{g}}lu}},\ }\bibfield  {title} {\emph
  {\bibinfo {title} {Quantum fuel with multilevel atomic coherence for
  ultrahigh specific work in a photonic carnot engine},\ }}\href
  {https://doi.org/10.1103/PhysRevE.93.012145} {\bibfield  {journal} {\bibinfo
  {journal} {Phys. Rev. E}\ }\textbf {\bibinfo {volume} {93}},\ \bibinfo
  {pages} {012145} (\bibinfo {year} {2016})}\BibitemShut {NoStop}%
\bibitem [{\citenamefont {Hardal}\ and\ \citenamefont
  {M{\"u}stecapl{\i}o{\u{g}}lu}(2015)}]{hardal2015superradiant}%
  \BibitemOpen
  \bibfield  {author} {\bibinfo {author} {\bibfnamefont {A.~{\"U}.~C.}\
  \bibnamefont {Hardal}}\ and\ \bibinfo {author} {\bibfnamefont {{\"O}.~E.}\
  \bibnamefont {M{\"u}stecapl{\i}o{\u{g}}lu}},\ }\bibfield  {title} {\emph
  {\bibinfo {title} {Superradiant quantum heat engine},\ }}\href
  {https://doi.org/10.1038/srep12953} {\bibfield  {journal} {\bibinfo
  {journal} {Sci. Rep.}\ }\textbf {\bibinfo {volume} {5}},\ \bibinfo {pages}
  {12953} (\bibinfo {year} {2015})}\BibitemShut {NoStop}%
\bibitem [{\citenamefont {Campisi}\ and\ \citenamefont
  {Fazio}(2016)}]{campisi_power_2016}%
  \BibitemOpen
  \bibfield  {author} {\bibinfo {author} {\bibfnamefont {M.}~\bibnamefont
  {Campisi}}\ and\ \bibinfo {author} {\bibfnamefont {R.}~\bibnamefont
  {Fazio}},\ }\bibfield  {title} {\emph {\bibinfo {title} {The power of a
  critical heat engine},\ }}\href {\doibase 10.1038/ncomms11895} {\bibfield
  {journal} {\bibinfo  {journal} {Nat. Comm.}\ }\textbf {\bibinfo {volume}
  {7}},\ \bibinfo {pages} {11895} (\bibinfo {year} {2016})}\BibitemShut
  {NoStop}%
\bibitem [{\citenamefont {Kosloff}\ and\ \citenamefont
  {Levy}(2014)}]{kosloff_quantum_2014}%
  \BibitemOpen
  \bibfield  {author} {\bibinfo {author} {\bibfnamefont {R.}~\bibnamefont
  {Kosloff}}\ and\ \bibinfo {author} {\bibfnamefont {A.}~\bibnamefont {Levy}},\
  }\bibfield  {title} {\emph {\bibinfo {title} {Quantum {Heat} {Engines} and
  {Refrigerators}: {Continuous} {Devices}},\ }}\href {\doibase
  10.1146/annurev-physchem-040513-103724} {\bibfield  {journal} {\bibinfo
  {journal} {Annu. Rev. Phys. Chem.}\ }\textbf {\bibinfo {volume} {65}},\
  \bibinfo {pages} {365} (\bibinfo {year} {2014})}\BibitemShut {NoStop}%
\bibitem [{\citenamefont {Tonner}\ and\ \citenamefont
  {Mahler}(2005)}]{tonner_autonomous_2005}%
  \BibitemOpen
  \bibfield  {author} {\bibinfo {author} {\bibfnamefont {F.}~\bibnamefont
  {Tonner}}\ and\ \bibinfo {author} {\bibfnamefont {G.}~\bibnamefont
  {Mahler}},\ }\bibfield  {title} {\emph {\bibinfo {title} {Autonomous quantum
  thermodynamic machines},\ }}\href {\doibase 10.1103/PhysRevE.72.066118}
  {\bibfield  {journal} {\bibinfo  {journal} {Phys. Rev. E}\ }\textbf {\bibinfo
  {volume} {72}},\ \bibinfo {pages} {066118} (\bibinfo {year}
  {2005})}\BibitemShut {NoStop}%
\bibitem [{\citenamefont {Anders}\ and\ \citenamefont
  {Giovannetti}(2013)}]{anders_thermodynamics_2013}%
  \BibitemOpen
  \bibfield  {author} {\bibinfo {author} {\bibfnamefont {J.}~\bibnamefont
  {Anders}}\ and\ \bibinfo {author} {\bibfnamefont {V.}~\bibnamefont
  {Giovannetti}},\ }\bibfield  {title} {\emph {\bibinfo {title} {Thermodynamics
  of discrete quantum processes},\ }}\href {\doibase
  10.1088/1367-2630/15/3/033022} {\bibfield  {journal} {\bibinfo  {journal}
  {New J. Phys.}\ }\textbf {\bibinfo {volume} {15}},\ \bibinfo {pages} {033022}
  (\bibinfo {year} {2013})}\BibitemShut {NoStop}%
\bibitem [{\citenamefont {Zhang}\ \emph
  {et~al.}(2014{\natexlab{a}})\citenamefont {Zhang}, \citenamefont {Bariani},\
  and\ \citenamefont {Meystre}}]{zhang_quantum_2014}%
  \BibitemOpen
  \bibfield  {author} {\bibinfo {author} {\bibfnamefont {K.}~\bibnamefont
  {Zhang}}, \bibinfo {author} {\bibfnamefont {F.}~\bibnamefont {Bariani}}, \
  and\ \bibinfo {author} {\bibfnamefont {P.}~\bibnamefont {Meystre}},\
  }\bibfield  {title} {\emph {\bibinfo {title} {Quantum {Optomechanical} {Heat}
  {Engine}},\ }}\href {\doibase 10.1103/PhysRevLett.112.150602} {\bibfield
  {journal} {\bibinfo  {journal} {Phys. Rev. Lett.}\ }\textbf {\bibinfo
  {volume} {112}},\ \bibinfo {pages} {150602} (\bibinfo {year}
  {2014}{\natexlab{a}})}\BibitemShut {NoStop}%
\bibitem [{\citenamefont {Zhang}\ \emph
  {et~al.}(2014{\natexlab{b}})\citenamefont {Zhang}, \citenamefont {Bariani},\
  and\ \citenamefont {Meystre}}]{zhang_theory_2014}%
  \BibitemOpen
  \bibfield  {author} {\bibinfo {author} {\bibfnamefont {K.}~\bibnamefont
  {Zhang}}, \bibinfo {author} {\bibfnamefont {F.}~\bibnamefont {Bariani}}, \
  and\ \bibinfo {author} {\bibfnamefont {P.}~\bibnamefont {Meystre}},\
  }\bibfield  {title} {\emph {\bibinfo {title} {Theory of an optomechanical
  quantum heat engine},\ }}\href {\doibase 10.1103/PhysRevA.90.023819}
  {\bibfield  {journal} {\bibinfo  {journal} {Phys. Rev. A}\ }\textbf {\bibinfo
  {volume} {90}},\ \bibinfo {pages} {023819} (\bibinfo {year}
  {2014}{\natexlab{b}})}\BibitemShut {NoStop}%
\bibitem [{\citenamefont {Zhang}\ and\ \citenamefont
  {Zhang}(2017)}]{zhang_quantum_2017}%
  \BibitemOpen
  \bibfield  {author} {\bibinfo {author} {\bibfnamefont {K.}~\bibnamefont
  {Zhang}}\ and\ \bibinfo {author} {\bibfnamefont {W.}~\bibnamefont {Zhang}},\
  }\bibfield  {title} {\emph {\bibinfo {title} {Quantum optomechanical
  straight-twin engine},\ }}\href {\doibase 10.1103/PhysRevA.95.053870}
  {\bibfield  {journal} {\bibinfo  {journal} {Phys. Rev. A}\ }\textbf {\bibinfo
  {volume} {95}},\ \bibinfo {pages} {053870} (\bibinfo {year}
  {2017})}\BibitemShut {NoStop}%
\bibitem [{\citenamefont {Mari}\ \emph {et~al.}(2015)\citenamefont {Mari},
  \citenamefont {Farace},\ and\ \citenamefont
  {Giovannetti}}]{mari_quantum_2015}%
  \BibitemOpen
  \bibfield  {author} {\bibinfo {author} {\bibfnamefont {A.}~\bibnamefont
  {Mari}}, \bibinfo {author} {\bibfnamefont {A.}~\bibnamefont {Farace}}, \ and\
  \bibinfo {author} {\bibfnamefont {V.}~\bibnamefont {Giovannetti}},\
  }\bibfield  {title} {\emph {\bibinfo {title} {Quantum optomechanical piston
  engines powered by heat},\ }}\href {\doibase 10.1088/0953-4075/48/17/175501}
  {\bibfield  {journal} {\bibinfo  {journal} {J. Phys. B At. Mol. Opt. Phys.}\
  }\textbf {\bibinfo {volume} {48}},\ \bibinfo {pages} {175501} (\bibinfo
  {year} {2015})}\BibitemShut {NoStop}%
\bibitem [{\citenamefont {Gelbwaser-Klimovsky}\ and\ \citenamefont
  {Kurizki}(2015)}]{gelbwaser-klimovsky_work_2015}%
  \BibitemOpen
  \bibfield  {author} {\bibinfo {author} {\bibfnamefont {D.}~\bibnamefont
  {Gelbwaser-Klimovsky}}\ and\ \bibinfo {author} {\bibfnamefont
  {G.}~\bibnamefont {Kurizki}},\ }\bibfield  {title} {\emph {\bibinfo {title}
  {Work extraction from heat-powered quantized optomechanical setups},\ }}\href
  {\doibase 10.1038/srep07809} {\bibfield  {journal} {\bibinfo  {journal} {Sci.
  Rep.}\ }\textbf {\bibinfo {volume} {5}},\ \bibinfo {pages} {7809} (\bibinfo
  {year} {2015})}\BibitemShut {NoStop}%
\bibitem [{\citenamefont {Ro{\ss}nagel}\ \emph {et~al.}(2014)\citenamefont
  {Ro{\ss}nagel}, \citenamefont {Abah}, \citenamefont {Schmidt-Kaler},
  \citenamefont {Singer},\ and\ \citenamefont
  {Lutz}}]{rosnagel_nanoscale_2014}%
  \BibitemOpen
  \bibfield  {author} {\bibinfo {author} {\bibfnamefont {J.}~\bibnamefont
  {Ro{\ss}nagel}}, \bibinfo {author} {\bibfnamefont {O.}~\bibnamefont {Abah}},
  \bibinfo {author} {\bibfnamefont {F.}~\bibnamefont {Schmidt-Kaler}}, \bibinfo
  {author} {\bibfnamefont {K.}~\bibnamefont {Singer}}, \ and\ \bibinfo {author}
  {\bibfnamefont {E.}~\bibnamefont {Lutz}},\ }\bibfield  {title} {\emph
  {\bibinfo {title} {Nanoscale {Heat} {Engine} {Beyond} the {Carnot} {Limit}},\
  }}\href {\doibase 10.1103/PhysRevLett.112.030602} {\bibfield  {journal}
  {\bibinfo  {journal} {Phys. Rev. Lett.}\ }\textbf {\bibinfo {volume} {112}},\
  \bibinfo {pages} {030602} (\bibinfo {year} {2014})}\BibitemShut {NoStop}%
\bibitem [{\citenamefont {Allahverdyan}\ and\ \citenamefont
  {Nieuwenhuizen}(2000)}]{allahverdyan_extraction_2000}%
  \BibitemOpen
  \bibfield  {author} {\bibinfo {author} {\bibfnamefont {A.~E.}\ \bibnamefont
  {Allahverdyan}}\ and\ \bibinfo {author} {\bibfnamefont {T.~M.}\ \bibnamefont
  {Nieuwenhuizen}},\ }\bibfield  {title} {\emph {\bibinfo {title} {Extraction
  of {Work} from a {Single} {Thermal} {Bath} in the {Quantum} {Regime}},\
  }}\href {\doibase 10.1103/PhysRevLett.85.1799} {\bibfield  {journal}
  {\bibinfo  {journal} {Phys. Rev. Lett.}\ }\textbf {\bibinfo {volume} {85}},\
  \bibinfo {pages} {1799} (\bibinfo {year} {2000})}\BibitemShut {NoStop}%
\bibitem [{\citenamefont {Korzekwa}\ \emph {et~al.}(2016)\citenamefont
  {Korzekwa}, \citenamefont {Lostaglio}, \citenamefont {Oppenheim},\ and\
  \citenamefont {Jennings}}]{korzekwa_extraction_2016}%
  \BibitemOpen
  \bibfield  {author} {\bibinfo {author} {\bibfnamefont {K.}~\bibnamefont
  {Korzekwa}}, \bibinfo {author} {\bibfnamefont {M.}~\bibnamefont {Lostaglio}},
  \bibinfo {author} {\bibfnamefont {J.}~\bibnamefont {Oppenheim}}, \ and\
  \bibinfo {author} {\bibfnamefont {D.}~\bibnamefont {Jennings}},\ }\bibfield
  {title} {\emph {\bibinfo {title} {The extraction of work from quantum
  coherence},\ }}\href {\doibase 10.1088/1367-2630/18/2/023045} {\bibfield
  {journal} {\bibinfo  {journal} {New J. Phys.}\ }\textbf {\bibinfo {volume}
  {18}},\ \bibinfo {pages} {023045} (\bibinfo {year} {2016})}\BibitemShut
  {NoStop}%
\bibitem [{\citenamefont {Perarnau-Llobet}\ \emph {et~al.}(2015)\citenamefont
  {Perarnau-Llobet}, \citenamefont {Hovhannisyan}, \citenamefont {Huber},
  \citenamefont {Skrzypczyk}, \citenamefont {Brunner},\ and\ \citenamefont
  {Ac{\'\i}n}}]{perarnau-llobet_extractable_2015}%
  \BibitemOpen
  \bibfield  {author} {\bibinfo {author} {\bibfnamefont {M.}~\bibnamefont
  {Perarnau-Llobet}}, \bibinfo {author} {\bibfnamefont {K.~V.}\ \bibnamefont
  {Hovhannisyan}}, \bibinfo {author} {\bibfnamefont {M.}~\bibnamefont {Huber}},
  \bibinfo {author} {\bibfnamefont {P.}~\bibnamefont {Skrzypczyk}}, \bibinfo
  {author} {\bibfnamefont {N.}~\bibnamefont {Brunner}}, \ and\ \bibinfo
  {author} {\bibfnamefont {A.}~\bibnamefont {Ac{\'\i}n}},\ }\bibfield  {title}
  {\emph {\bibinfo {title} {Extractable {Work} from {Correlations}},\ }}\href
  {\doibase 10.1103/PhysRevX.5.041011} {\bibfield  {journal} {\bibinfo
  {journal} {Phys. Rev. X}\ }\textbf {\bibinfo {volume} {5}},\ \bibinfo {pages}
  {041011} (\bibinfo {year} {2015})}\BibitemShut {NoStop}%
\bibitem [{\citenamefont {Plastina}\ \emph {et~al.}(2014)\citenamefont
  {Plastina}, \citenamefont {Alecce}, \citenamefont {Apollaro}, \citenamefont
  {Falcone}, \citenamefont {Francica}, \citenamefont {Galve}, \citenamefont
  {Lo~Gullo},\ and\ \citenamefont {Zambrini}}]{plastina_irreversible_2014}%
  \BibitemOpen
  \bibfield  {author} {\bibinfo {author} {\bibfnamefont {F.}~\bibnamefont
  {Plastina}}, \bibinfo {author} {\bibfnamefont {A.}~\bibnamefont {Alecce}},
  \bibinfo {author} {\bibfnamefont {T.}~\bibnamefont {Apollaro}}, \bibinfo
  {author} {\bibfnamefont {G.}~\bibnamefont {Falcone}}, \bibinfo {author}
  {\bibfnamefont {G.}~\bibnamefont {Francica}}, \bibinfo {author}
  {\bibfnamefont {F.}~\bibnamefont {Galve}}, \bibinfo {author} {\bibfnamefont
  {N.}~\bibnamefont {Lo~Gullo}}, \ and\ \bibinfo {author} {\bibfnamefont
  {R.}~\bibnamefont {Zambrini}},\ }\bibfield  {title} {\emph {\bibinfo {title}
  {Irreversible {Work} and {Inner} {Friction} in {Quantum} {Thermodynamic}
  {Processes}},\ }}\href {\doibase 10.1103/PhysRevLett.113.260601} {\bibfield
  {journal} {\bibinfo  {journal} {Phys. Rev. Lett.}\ }\textbf {\bibinfo
  {volume} {113}},\ \bibinfo {pages} {260601} (\bibinfo {year}
  {2014})}\BibitemShut {NoStop}%
\bibitem [{\citenamefont {Campo}\ \emph {et~al.}(2014)\citenamefont {Campo},
  \citenamefont {Goold},\ and\ \citenamefont {Paternostro}}]{campo_more_2014}%
  \BibitemOpen
  \bibfield  {author} {\bibinfo {author} {\bibfnamefont {A.~d.}\ \bibnamefont
  {Campo}}, \bibinfo {author} {\bibfnamefont {J.}~\bibnamefont {Goold}}, \ and\
  \bibinfo {author} {\bibfnamefont {M.}~\bibnamefont {Paternostro}},\
  }\bibfield  {title} {\emph {\bibinfo {title} {More bang for your buck:
  {Super}-adiabatic quantum engines},\ }}\href {\doibase 10.1038/srep06208}
  {\bibfield  {journal} {\bibinfo  {journal} {Sci. Rep.}\ }\textbf {\bibinfo
  {volume} {4}},\ \bibinfo {pages} {6208} (\bibinfo {year} {2014})}\BibitemShut
  {NoStop}%
\bibitem [{\citenamefont {Roulet}\ \emph {et~al.}(2017)\citenamefont {Roulet},
  \citenamefont {Nimmrichter}, \citenamefont {Arrazola}, \citenamefont {Seah},\
  and\ \citenamefont {Scarani}}]{roulet_autonomous_2017}%
  \BibitemOpen
  \bibfield  {author} {\bibinfo {author} {\bibfnamefont {A.}~\bibnamefont
  {Roulet}}, \bibinfo {author} {\bibfnamefont {S.}~\bibnamefont {Nimmrichter}},
  \bibinfo {author} {\bibfnamefont {J.~M.}\ \bibnamefont {Arrazola}}, \bibinfo
  {author} {\bibfnamefont {S.}~\bibnamefont {Seah}}, \ and\ \bibinfo {author}
  {\bibfnamefont {V.}~\bibnamefont {Scarani}},\ }\bibfield  {title} {\emph
  {\bibinfo {title} {Autonomous rotor heat engine},\ }}\href {\doibase
  10.1103/PhysRevE.95.062131} {\bibfield  {journal} {\bibinfo  {journal} {Phys.
  Rev. E}\ }\textbf {\bibinfo {volume} {95}},\ \bibinfo {pages} {062131}
  (\bibinfo {year} {2017})}\BibitemShut {NoStop}%
\bibitem [{\citenamefont {Hofer}\ \emph
  {et~al.}(2016{\natexlab{a}})\citenamefont {Hofer}, \citenamefont {Souquet},\
  and\ \citenamefont {Clerk}}]{hofer2016quantum}%
  \BibitemOpen
  \bibfield  {author} {\bibinfo {author} {\bibfnamefont {P.~P.}\ \bibnamefont
  {Hofer}}, \bibinfo {author} {\bibfnamefont {J.-R.}\ \bibnamefont {Souquet}},
  \ and\ \bibinfo {author} {\bibfnamefont {A.~A.}\ \bibnamefont {Clerk}},\
  }\bibfield  {title} {\emph {\bibinfo {title} {Quantum heat engine based on
  photon-assisted cooper pair tunneling},\ }}\href
  {https://doi.org/10.1103/PhysRevB.93.041418} {\bibfield  {journal} {\bibinfo
  {journal} {Physical Review B}\ }\textbf {\bibinfo {volume} {93}},\ \bibinfo
  {pages} {041418} (\bibinfo {year} {2016}{\natexlab{a}})}\BibitemShut
  {NoStop}%
\bibitem [{\citenamefont {Hofer}\ \emph
  {et~al.}(2017{\natexlab{a}})\citenamefont {Hofer}, \citenamefont {Brask},
  \citenamefont {Perarnau-Llobet},\ and\ \citenamefont
  {Brunner}}]{hofer2017quantum}%
  \BibitemOpen
  \bibfield  {author} {\bibinfo {author} {\bibfnamefont {P.~P.}\ \bibnamefont
  {Hofer}}, \bibinfo {author} {\bibfnamefont {J.~B.}\ \bibnamefont {Brask}},
  \bibinfo {author} {\bibfnamefont {M.}~\bibnamefont {Perarnau-Llobet}}, \ and\
  \bibinfo {author} {\bibfnamefont {N.}~\bibnamefont {Brunner}},\ }\bibfield
  {title} {\emph {\bibinfo {title} {Quantum thermal machine as a thermometer},\
  }}\href {https://arxiv.org/abs/1703.03719} {\bibfield  {journal} {\bibinfo
  {journal} {arXiv preprint arXiv:1703.03719}\ } (\bibinfo {year}
  {2017}{\natexlab{a}})}\BibitemShut {NoStop}%
\bibitem [{\citenamefont {Hofer}\ \emph
  {et~al.}(2016{\natexlab{b}})\citenamefont {Hofer}, \citenamefont
  {Perarnau-Llobet}, \citenamefont {Brask}, \citenamefont {Silva},
  \citenamefont {Huber},\ and\ \citenamefont {Brunner}}]{hofer2016autonomous}%
  \BibitemOpen
  \bibfield  {author} {\bibinfo {author} {\bibfnamefont {P.~P.}\ \bibnamefont
  {Hofer}}, \bibinfo {author} {\bibfnamefont {M.}~\bibnamefont
  {Perarnau-Llobet}}, \bibinfo {author} {\bibfnamefont {J.~B.}\ \bibnamefont
  {Brask}}, \bibinfo {author} {\bibfnamefont {R.}~\bibnamefont {Silva}},
  \bibinfo {author} {\bibfnamefont {M.}~\bibnamefont {Huber}}, \ and\ \bibinfo
  {author} {\bibfnamefont {N.}~\bibnamefont {Brunner}},\ }\bibfield  {title}
  {\emph {\bibinfo {title} {Autonomous quantum refrigerator in a circuit qed
  architecture based on a josephson junction},\ }}\href
  {https://doi.org/10.1103/PhysRevB.94.235420} {\bibfield  {journal} {\bibinfo
  {journal} {Physical Review B}\ }\textbf {\bibinfo {volume} {94}},\ \bibinfo
  {pages} {235420} (\bibinfo {year} {2016}{\natexlab{b}})}\BibitemShut
  {NoStop}%
\bibitem [{\citenamefont {Karimi}\ and\ \citenamefont
  {Pekola}(2016)}]{karimi2016otto}%
  \BibitemOpen
  \bibfield  {author} {\bibinfo {author} {\bibfnamefont {B.}~\bibnamefont
  {Karimi}}\ and\ \bibinfo {author} {\bibfnamefont {J.}~\bibnamefont
  {Pekola}},\ }\bibfield  {title} {\emph {\bibinfo {title} {Otto refrigerator
  based on a superconducting qubit: Classical and quantum performance},\
  }}\href {https://doi.org/10.1103/PhysRevB.94.184503} {\bibfield  {journal}
  {\bibinfo  {journal} {Physical Review B}\ }\textbf {\bibinfo {volume} {94}},\
  \bibinfo {pages} {184503} (\bibinfo {year} {2016})}\BibitemShut {NoStop}%
\bibitem [{\citenamefont {Ro{\ss}nagel}\ \emph {et~al.}(2016)\citenamefont
  {Ro{\ss}nagel}, \citenamefont {Dawkins}, \citenamefont {Tolazzi},
  \citenamefont {Abah}, \citenamefont {Lutz}, \citenamefont {Schmidt-Kaler},\
  and\ \citenamefont {Singer}}]{rosnagel_single-atom_2016}%
  \BibitemOpen
  \bibfield  {author} {\bibinfo {author} {\bibfnamefont {J.}~\bibnamefont
  {Ro{\ss}nagel}}, \bibinfo {author} {\bibfnamefont {S.~T.}\ \bibnamefont
  {Dawkins}}, \bibinfo {author} {\bibfnamefont {K.~N.}\ \bibnamefont
  {Tolazzi}}, \bibinfo {author} {\bibfnamefont {O.}~\bibnamefont {Abah}},
  \bibinfo {author} {\bibfnamefont {E.}~\bibnamefont {Lutz}}, \bibinfo {author}
  {\bibfnamefont {F.}~\bibnamefont {Schmidt-Kaler}}, \ and\ \bibinfo {author}
  {\bibfnamefont {K.}~\bibnamefont {Singer}},\ }\bibfield  {title} {\emph
  {\bibinfo {title} {A single-atom heat engine},\ }}\href {\doibase
  10.1126/science.aad6320} {\bibfield  {journal} {\bibinfo  {journal}
  {Science}\ }\textbf {\bibinfo {volume} {352}},\ \bibinfo {pages} {325}
  (\bibinfo {year} {2016})}\BibitemShut {NoStop}%
\bibitem [{\citenamefont {Uzdin}\ \emph {et~al.}(2015)\citenamefont {Uzdin},
  \citenamefont {Levy},\ and\ \citenamefont
  {Kosloff}}]{uzdin_equivalence_2015}%
  \BibitemOpen
  \bibfield  {author} {\bibinfo {author} {\bibfnamefont {R.}~\bibnamefont
  {Uzdin}}, \bibinfo {author} {\bibfnamefont {A.}~\bibnamefont {Levy}}, \ and\
  \bibinfo {author} {\bibfnamefont {R.}~\bibnamefont {Kosloff}},\ }\bibfield
  {title} {\emph {\bibinfo {title} {Equivalence of {Quantum} {Heat} {Machines},
  and {Quantum}-{Thermodynamic} {Signatures}},\ }}\href {\doibase
  10.1103/PhysRevX.5.031044} {\bibfield  {journal} {\bibinfo  {journal} {Phys.
  Rev. X}\ }\textbf {\bibinfo {volume} {5}},\ \bibinfo {pages} {031044}
  (\bibinfo {year} {2015})}\BibitemShut {NoStop}%
\bibitem [{\citenamefont {Mukherjee}\ \emph {et~al.}(2016)\citenamefont
  {Mukherjee}, \citenamefont {Niedenzu}, \citenamefont {Kofman},\ and\
  \citenamefont {Kurizki}}]{mukherjee_speed_2016}%
  \BibitemOpen
  \bibfield  {author} {\bibinfo {author} {\bibfnamefont {V.}~\bibnamefont
  {Mukherjee}}, \bibinfo {author} {\bibfnamefont {W.}~\bibnamefont {Niedenzu}},
  \bibinfo {author} {\bibfnamefont {A.~G.}\ \bibnamefont {Kofman}}, \ and\
  \bibinfo {author} {\bibfnamefont {G.}~\bibnamefont {Kurizki}},\ }\bibfield
  {title} {\emph {\bibinfo {title} {Speed and {Efficiency} {Limits} of
  {Multilevel} {Incoherent} {Heat} {Engines}},\ }}\href
  {http://arxiv.org/abs/1607.08452} {\bibfield  {journal} {\bibinfo  {journal}
  {arXiv:1607.08452 [cond-mat, physics:quant-ph]}\ } (\bibinfo {year}
  {2016})},\ \bibinfo {note} {arXiv: 1607.08452}\BibitemShut {NoStop}%
\bibitem [{\citenamefont {Quan}\ \emph {et~al.}(2006)\citenamefont {Quan},
  \citenamefont {Zhang},\ and\ \citenamefont
  {Sun}}]{quan_quantum-classical_2006}%
  \BibitemOpen
  \bibfield  {author} {\bibinfo {author} {\bibfnamefont {H.~T.}\ \bibnamefont
  {Quan}}, \bibinfo {author} {\bibfnamefont {P.}~\bibnamefont {Zhang}}, \ and\
  \bibinfo {author} {\bibfnamefont {C.~P.}\ \bibnamefont {Sun}},\ }\bibfield
  {title} {\emph {\bibinfo {title} {Quantum-classical transition of
  photon-{Carnot} engine induced by quantum decoherence},\ }}\href {\doibase
  10.1103/PhysRevE.73.036122} {\bibfield  {journal} {\bibinfo  {journal} {Phys.
  Rev. E}\ }\textbf {\bibinfo {volume} {73}},\ \bibinfo {pages} {036122}
  (\bibinfo {year} {2006})}\BibitemShut {NoStop}%
\bibitem [{\citenamefont {Johansson}\ \emph {et~al.}(2014)\citenamefont
  {Johansson}, \citenamefont {Johansson},\ and\ \citenamefont
  {Nori}}]{johansson_optomechanical-like_2014}%
  \BibitemOpen
  \bibfield  {author} {\bibinfo {author} {\bibfnamefont {J.~R.}\ \bibnamefont
  {Johansson}}, \bibinfo {author} {\bibfnamefont {G.}~\bibnamefont
  {Johansson}}, \ and\ \bibinfo {author} {\bibfnamefont {F.}~\bibnamefont
  {Nori}},\ }\bibfield  {title} {\emph {\bibinfo {title} {Optomechanical-like
  coupling between superconducting resonators},\ }}\href {\doibase
  10.1103/PhysRevA.90.053833} {\bibfield  {journal} {\bibinfo  {journal} {Phys.
  Rev. A}\ }\textbf {\bibinfo {volume} {90}},\ \bibinfo {pages} {053833}
  (\bibinfo {year} {2014})}\BibitemShut {NoStop}%
\bibitem [{\citenamefont {Holland}\ \emph {et~al.}(1990)\citenamefont
  {Holland}, \citenamefont {Collett}, \citenamefont {Walls},\ and\
  \citenamefont {Levenson}}]{holland1990nonideal}%
  \BibitemOpen
  \bibfield  {author} {\bibinfo {author} {\bibfnamefont {M.}~\bibnamefont
  {Holland}}, \bibinfo {author} {\bibfnamefont {M.}~\bibnamefont {Collett}},
  \bibinfo {author} {\bibfnamefont {D.}~\bibnamefont {Walls}}, \ and\ \bibinfo
  {author} {\bibfnamefont {M.}~\bibnamefont {Levenson}},\ }\bibfield  {title}
  {\emph {\bibinfo {title} {Nonideal quantum nondemolition measurements},\
  }}\href {https://doi.org/10.1103/PhysRevA.42.2995} {\bibfield  {journal}
  {\bibinfo  {journal} {Phys. Rev. A}\ }\textbf {\bibinfo {volume} {42}},\
  \bibinfo {pages} {2995} (\bibinfo {year} {1990})}\BibitemShut {NoStop}%
\bibitem [{\citenamefont {Mancini}\ \emph {et~al.}(1997)\citenamefont
  {Mancini}, \citenamefont {Man'ko},\ and\ \citenamefont
  {Tombesi}}]{mancini1997ponderomotive}%
  \BibitemOpen
  \bibfield  {author} {\bibinfo {author} {\bibfnamefont {S.}~\bibnamefont
  {Mancini}}, \bibinfo {author} {\bibfnamefont {V.}~\bibnamefont {Man'ko}}, \
  and\ \bibinfo {author} {\bibfnamefont {P.}~\bibnamefont {Tombesi}},\
  }\bibfield  {title} {\emph {\bibinfo {title} {Ponderomotive control of
  quantum macroscopic coherence},\ }}\href
  {https://doi.org/10.1103/PhysRevA.55.3042} {\bibfield  {journal} {\bibinfo
  {journal} {Phys. Rev. A}\ }\textbf {\bibinfo {volume} {55}},\ \bibinfo
  {pages} {3042} (\bibinfo {year} {1997})}\BibitemShut {NoStop}%
\bibitem [{\citenamefont {Fink}\ \emph {et~al.}(2010)\citenamefont {Fink},
  \citenamefont {Steffen}, \citenamefont {Studer}, \citenamefont {Bishop},
  \citenamefont {Baur}, \citenamefont {Bianchetti}, \citenamefont {Bozyigit},
  \citenamefont {Lang}, \citenamefont {Filipp}, \citenamefont {Leek},\ and\
  \citenamefont {Wallraff}}]{fink_quantum-classical_2010}%
  \BibitemOpen
  \bibfield  {author} {\bibinfo {author} {\bibfnamefont {J.~M.}\ \bibnamefont
  {Fink}}, \bibinfo {author} {\bibfnamefont {L.}~\bibnamefont {Steffen}},
  \bibinfo {author} {\bibfnamefont {P.}~\bibnamefont {Studer}}, \bibinfo
  {author} {\bibfnamefont {L.~S.}\ \bibnamefont {Bishop}}, \bibinfo {author}
  {\bibfnamefont {M.}~\bibnamefont {Baur}}, \bibinfo {author} {\bibfnamefont
  {R.}~\bibnamefont {Bianchetti}}, \bibinfo {author} {\bibfnamefont
  {D.}~\bibnamefont {Bozyigit}}, \bibinfo {author} {\bibfnamefont
  {C.}~\bibnamefont {Lang}}, \bibinfo {author} {\bibfnamefont {S.}~\bibnamefont
  {Filipp}}, \bibinfo {author} {\bibfnamefont {P.~J.}\ \bibnamefont {Leek}}, \
  and\ \bibinfo {author} {\bibfnamefont {A.}~\bibnamefont {Wallraff}},\
  }\bibfield  {title} {\emph {\bibinfo {title} {Quantum-{To}-{Classical}
  {Transition} in {Cavity} {Quantum} {Electrodynamics}},\ }}\href {\doibase
  10.1103/PhysRevLett.105.163601} {\bibfield  {journal} {\bibinfo  {journal}
  {Phys. Rev. Lett.}\ }\textbf {\bibinfo {volume} {105}},\ \bibinfo {pages}
  {163601} (\bibinfo {year} {2010})}\BibitemShut {NoStop}%
\bibitem [{\citenamefont {Levy}\ and\ \citenamefont
  {Kosloff}(2014)}]{levy_local_2014}%
  \BibitemOpen
  \bibfield  {author} {\bibinfo {author} {\bibfnamefont {A.}~\bibnamefont
  {Levy}}\ and\ \bibinfo {author} {\bibfnamefont {R.}~\bibnamefont {Kosloff}},\
  }\bibfield  {title} {\emph {\bibinfo {title} {The local approach to quantum
  transport may violate the second law of thermodynamics},\ }}\href {\doibase
  10.1209/0295-5075/107/20004} {\bibfield  {journal} {\bibinfo  {journal}
  {Europhys. Lett.}\ }\textbf {\bibinfo {volume} {107}},\ \bibinfo {pages}
  {20004} (\bibinfo {year} {2014})}\BibitemShut {NoStop}%
\bibitem [{\citenamefont {Hofer}\ \emph
  {et~al.}(2017{\natexlab{b}})\citenamefont {Hofer}, \citenamefont
  {Perarnau-Llobet}, \citenamefont {Miranda}, \citenamefont {Haack},
  \citenamefont {Silva}, \citenamefont {Brask},\ and\ \citenamefont
  {Brunner}}]{hofer2017markovian}%
  \BibitemOpen
  \bibfield  {author} {\bibinfo {author} {\bibfnamefont {P.~P.}\ \bibnamefont
  {Hofer}}, \bibinfo {author} {\bibfnamefont {M.}~\bibnamefont
  {Perarnau-Llobet}}, \bibinfo {author} {\bibfnamefont {L.~D.~M.}\ \bibnamefont
  {Miranda}}, \bibinfo {author} {\bibfnamefont {G.}~\bibnamefont {Haack}},
  \bibinfo {author} {\bibfnamefont {R.}~\bibnamefont {Silva}}, \bibinfo
  {author} {\bibfnamefont {J.~B.}\ \bibnamefont {Brask}}, \ and\ \bibinfo
  {author} {\bibfnamefont {N.}~\bibnamefont {Brunner}},\ }\bibfield  {title}
  {\emph {\bibinfo {title} {Markovian master equations for quantum thermal
  machines: local vs global approach},\ }}\href
  {https://arxiv.org/abs/1707.09211} {\bibfield  {journal} {\bibinfo  {journal}
  {arXiv preprint arXiv:1707.09211}\ } (\bibinfo {year}
  {2017}{\natexlab{b}})}\BibitemShut {NoStop}%
\bibitem [{\citenamefont {Gonz{\'a}lez}\ \emph {et~al.}(2017)\citenamefont
  {Gonz{\'a}lez}, \citenamefont {Correa}, \citenamefont {Nocerino},
  \citenamefont {Palao}, \citenamefont {Alonso},\ and\ \citenamefont
  {Adesso}}]{gonzalez2017testing}%
  \BibitemOpen
  \bibfield  {author} {\bibinfo {author} {\bibfnamefont {J.~O.}\ \bibnamefont
  {Gonz{\'a}lez}}, \bibinfo {author} {\bibfnamefont {L.~A.}\ \bibnamefont
  {Correa}}, \bibinfo {author} {\bibfnamefont {G.}~\bibnamefont {Nocerino}},
  \bibinfo {author} {\bibfnamefont {J.~P.}\ \bibnamefont {Palao}}, \bibinfo
  {author} {\bibfnamefont {D.}~\bibnamefont {Alonso}}, \ and\ \bibinfo {author}
  {\bibfnamefont {G.}~\bibnamefont {Adesso}},\ }\bibfield  {title} {\emph
  {\bibinfo {title} {Testing the validity of the local and global gkls master
  equations on an exactly solvable model},\ }}\href
  {https://arxiv.org/abs/1707.09228} {\bibfield  {journal} {\bibinfo  {journal}
  {arXiv preprint arXiv:1707.09228}\ } (\bibinfo {year} {2017})}\BibitemShut
  {NoStop}%
\bibitem [{\citenamefont {Hu}\ \emph {et~al.}(2015)\citenamefont {Hu},
  \citenamefont {Huang}, \citenamefont {Liao}, \citenamefont {Tian},\ and\
  \citenamefont {Goan}}]{hu_quantum_2015}%
  \BibitemOpen
  \bibfield  {author} {\bibinfo {author} {\bibfnamefont {D.}~\bibnamefont
  {Hu}}, \bibinfo {author} {\bibfnamefont {S.-Y.}\ \bibnamefont {Huang}},
  \bibinfo {author} {\bibfnamefont {J.-Q.}\ \bibnamefont {Liao}}, \bibinfo
  {author} {\bibfnamefont {L.}~\bibnamefont {Tian}}, \ and\ \bibinfo {author}
  {\bibfnamefont {H.-S.}\ \bibnamefont {Goan}},\ }\bibfield  {title} {\emph
  {\bibinfo {title} {Quantum coherence in ultrastrong optomechanics},\ }}\href
  {\doibase 10.1103/PhysRevA.91.013812} {\bibfield  {journal} {\bibinfo
  {journal} {Phys. Rev. A}\ }\textbf {\bibinfo {volume} {91}},\ \bibinfo
  {pages} {013812} (\bibinfo {year} {2015})}\BibitemShut {NoStop}%
\bibitem [{\citenamefont {Teufel}\ \emph {et~al.}()\citenamefont {Teufel},
  \citenamefont {Donner}, \citenamefont {Li}, \citenamefont {Harlow},
  \citenamefont {Allman}, \citenamefont {Cicak}, \citenamefont {Sirois},
  \citenamefont {Whittaker}, \citenamefont {Lehnert},\ and\ \citenamefont
  {Simmonds}}]{teufel2011sideband}%
  \BibitemOpen
  \bibfield  {author} {\bibinfo {author} {\bibfnamefont {J.}~\bibnamefont
  {Teufel}}, \bibinfo {author} {\bibfnamefont {T.}~\bibnamefont {Donner}},
  \bibinfo {author} {\bibfnamefont {D.}~\bibnamefont {Li}}, \bibinfo {author}
  {\bibfnamefont {J.}~\bibnamefont {Harlow}}, \bibinfo {author} {\bibfnamefont
  {M.}~\bibnamefont {Allman}}, \bibinfo {author} {\bibfnamefont
  {K.}~\bibnamefont {Cicak}}, \bibinfo {author} {\bibfnamefont
  {A.}~\bibnamefont {Sirois}}, \bibinfo {author} {\bibfnamefont {J.~D.}\
  \bibnamefont {Whittaker}}, \bibinfo {author} {\bibfnamefont {K.}~\bibnamefont
  {Lehnert}}, \ and\ \bibinfo {author} {\bibfnamefont {R.~W.}\ \bibnamefont
  {Simmonds}},\ }\bibfield  {title} {\emph {\bibinfo {title} {Sideband cooling
  of micromechanical motion to the quantum ground state},\ }}\href
  {https://doi.org/10.1038/nature10261} {\bibfield  {journal} {\bibinfo
  {journal} {Nature}\ }\textbf {\bibinfo {volume} {475}},\ \bibinfo {pages}
  {359}}\BibitemShut {NoStop}%
\bibitem [{\citenamefont {Johansson}\ \emph {et~al.}(2013)\citenamefont
  {Johansson}, \citenamefont {Nation},\ and\ \citenamefont
  {Nori}}]{johansson2013qutip}%
  \BibitemOpen
  \bibfield  {author} {\bibinfo {author} {\bibfnamefont {J.}~\bibnamefont
  {Johansson}}, \bibinfo {author} {\bibfnamefont {P.}~\bibnamefont {Nation}}, \
  and\ \bibinfo {author} {\bibfnamefont {F.}~\bibnamefont {Nori}},\ }\bibfield
  {title} {\emph {\bibinfo {title} {Qutip 2: A python framework for the
  dynamics of open quantum systems},\ }}\href
  {https://doi.org/10.1016/j.cpc.2012.02.021} {\bibfield  {journal} {\bibinfo
  {journal} {Comput. Phys. Commun.}\ }\textbf {\bibinfo {volume} {184}},\
  \bibinfo {pages} {1234} (\bibinfo {year} {2013})}\BibitemShut {NoStop}%
\bibitem [{\citenamefont {Feldmann}\ and\ \citenamefont
  {Kosloff}(2004)}]{feldmann_characteristics_2004}%
  \BibitemOpen
  \bibfield  {author} {\bibinfo {author} {\bibfnamefont {T.}~\bibnamefont
  {Feldmann}}\ and\ \bibinfo {author} {\bibfnamefont {R.}~\bibnamefont
  {Kosloff}},\ }\bibfield  {title} {\emph {\bibinfo {title} {Characteristics of
  the limit cycle of a reciprocating quantum heat engine},\ }}\href {\doibase
  10.1103/PhysRevE.70.046110} {\bibfield  {journal} {\bibinfo  {journal} {Phys.
  Rev. E}\ }\textbf {\bibinfo {volume} {70}},\ \bibinfo {pages} {046110}
  (\bibinfo {year} {2004})}\BibitemShut {NoStop}%
\bibitem [{\citenamefont {da~Silva}\ \emph {et~al.}(2010)\citenamefont
  {da~Silva}, \citenamefont {Bozyigit}, \citenamefont {Wallraff},\ and\
  \citenamefont {Blais}}]{da_silva_schemes_2010}%
  \BibitemOpen
  \bibfield  {author} {\bibinfo {author} {\bibfnamefont {M.~P.}\ \bibnamefont
  {da~Silva}}, \bibinfo {author} {\bibfnamefont {D.}~\bibnamefont {Bozyigit}},
  \bibinfo {author} {\bibfnamefont {A.}~\bibnamefont {Wallraff}}, \ and\
  \bibinfo {author} {\bibfnamefont {A.}~\bibnamefont {Blais}},\ }\bibfield
  {title} {\emph {\bibinfo {title} {Schemes for the observation of photon
  correlation functions in circuit {QED} with linear detectors},\ }}\href
  {\doibase 10.1103/PhysRevA.82.043804} {\bibfield  {journal} {\bibinfo
  {journal} {Phys. Rev. A}\ }\textbf {\bibinfo {volume} {82}},\ \bibinfo
  {pages} {043804} (\bibinfo {year} {2010})}\BibitemShut {NoStop}%
\bibitem [{\citenamefont {Klaers}\ \emph {et~al.}(2017)\citenamefont {Klaers},
  \citenamefont {Faelt}, \citenamefont {Imamoglu},\ and\ \citenamefont
  {Togan}}]{klaers_squeezed_2017}%
  \BibitemOpen
  \bibfield  {author} {\bibinfo {author} {\bibfnamefont {J.}~\bibnamefont
  {Klaers}}, \bibinfo {author} {\bibfnamefont {S.}~\bibnamefont {Faelt}},
  \bibinfo {author} {\bibfnamefont {A.}~\bibnamefont {Imamoglu}}, \ and\
  \bibinfo {author} {\bibfnamefont {E.}~\bibnamefont {Togan}},\ }\bibfield
  {title} {\emph {\bibinfo {title} {Squeezed thermal reservoirs as a resource
  for a nano-mechanical engine beyond the {Carnot} limit},\ }}\href
  {http://arxiv.org/abs/1703.10024} {\bibfield  {journal} {\bibinfo  {journal}
  {arXiv:1703.10024 [cond-mat]}\ } (\bibinfo {year} {2017})}\BibitemShut
  {NoStop}%
\bibitem [{\citenamefont {Kubo}(1966)}]{kubo1966fluctuation}%
  \BibitemOpen
  \bibfield  {author} {\bibinfo {author} {\bibfnamefont {R.}~\bibnamefont
  {Kubo}},\ }\bibfield  {title} {\emph {\bibinfo {title} {The
  fluctuation-dissipation theorem},\ }}\href
  {http://stacks.iop.org/0034-4885/29/i=1/a=306} {\bibfield  {journal}
  {\bibinfo  {journal} {Rep. Prog. Phys.}\ }\textbf {\bibinfo {volume} {29}},\
  \bibinfo {pages} {255} (\bibinfo {year} {1966})}\BibitemShut {NoStop}%
\bibitem [{\citenamefont {Bern{\'a}d}\ and\ \citenamefont
  {Torres}(2015)}]{bernad_partly_2015}%
  \BibitemOpen
  \bibfield  {author} {\bibinfo {author} {\bibfnamefont {J.~Z.}\ \bibnamefont
  {Bern{\'a}d}}\ and\ \bibinfo {author} {\bibfnamefont {J.~M.}\ \bibnamefont
  {Torres}},\ }\bibfield  {title} {\emph {\bibinfo {title} {Partly invariant
  steady state of two interacting open quantum systems},\ }}\href {\doibase
  10.1103/PhysRevA.92.062114} {\bibfield  {journal} {\bibinfo  {journal} {Phys.
  Rev. A}\ }\textbf {\bibinfo {volume} {92}},\ \bibinfo {pages} {062114}
  (\bibinfo {year} {2015})}\BibitemShut {NoStop}%
\end{thebibliography}%
\end{document}